\documentclass[%
 reprint,
groupedaddress,
 amsmath,amssymb,
 aps,
 pra,
]{revtex4-2}

\usepackage{dcolumn}
\usepackage{bm}
\usepackage{mathtools} 
\usepackage{color}  
\usepackage{float}
\usepackage{braket}
\usepackage{mhchem}
\usepackage{bm}
\usepackage{bbold}
\usepackage{mathtools}
\usepackage{mhchem}

\usepackage[colorlinks=true,allcolors=blue]{hyperref}

\usepackage[caption=false, justification=centerlast]{subfig}
\usepackage[export]{adjustbox}
\usepackage{xcolor}           

\usepackage[normalem]{ulem}   
\newif\ifshowstrikes
\showstrikestrue  



\definecolor{mscolor}{rgb}{0,0.5,0.5}

\definecolor{AkScolor}{rgb}{0.4,0,0}

\definecolor{tgcolor}{rgb}{0.5,0,0.5}

{}
\definecolor{phcolor}{rgb}{0.5,0,0.5}

\newcommand{\UWM}{Department of Physics, University of Wisconsin-Madison, 1150 University Avenue, Madison, WI, 53706, USA}

\newcommand{\rsub}[1]{\textcolor{black}{#1}}

\newcommand \be{\begin{equation}}
\newcommand \ee{\end{equation}}
\newcommand \bea{\begin{eqnarray}}
\newcommand \eea{\end{eqnarray}}
\newcommand \bse{\begin{subequations}}
\newcommand \ese{\end{subequations}}

\begin{document}

\title{Exact steady state of perturbed open quantum systems }

\author{Omar Nagib}
 \email{onagib@wisc.edu}
\affiliation{\UWM}

\author{T. G. Walker}
\affiliation{\UWM}

\date{\today}

\begin{abstract}
 \rsub{We present a general non-perturbative method to determine the exact steady state of open quantum systems under perturbation}. \rsub{The method works for systems with a unique steady state and} the perturbation may be time-independent or periodic, and of arbitrarily large amplitude. Using \rsub{the Drazin inverse and \rsub{a single} diagonalization}, we construct an operator that generates the entire dependence of the steady state on the perturbation parameter. The approach also enables exact analytic operations—such as differentiation, integration, and ensemble averaging—with respect to the parameter, even when the steady state is computed numerically. We apply the method to three non-trivial open quantum systems, showing that it achieves exact results, with a computational speedup of \rsub{one to several orders of magnitude} for calculations requiring large sampling, \rsub{compared to previous approaches.}

\end{abstract}

\maketitle

\section{Introduction}
 
Many quantum processes, such as photochemistry, energy transport, quantum optics, electronic and spin resonance, quantum computing and sensing, are described by a quantum master equation \cite{QuantumMasterPro}. Finding the steady state of such open quantum systems is a problem of great theoretical and practical interest. For sufficiently large and complicated systems, only numerical solutions are possible. Consequently, many numerical approaches have been developed, including eigenvalue methods and $\rm LU$ decomposition \cite{Review_2015}, variational principles \cite{Variational_2015, VPT2025}, neural networks \cite{Neural_2019}, quantum trajectories \cite{Quantum_trajectory_1992,Trajectory_static_disorder_2010}, matrix product density operators \cite{MPDO}, and iterative methods \cite{Iter_2015}, to name a few. \rsub{For further theoretical and numerical methods on non-equilibrium steady states refer to Refs.  \cite{HEOM_2021,Entropy_2011,NESS_mesoscopic_2012,NESS_quantum_thermo_2015,Topp_2015,Ness_2017}, and for a general review of numerical techniques for the master equation see Refs. \cite{QuantumMasterPro, Review_2015, Many_body_review_2021} and the references within}. Numerical tools for efficient simulation of open quantum systems have also been created, such as QuTiP \cite{QuTiP}, HOQST \cite{HOQST}, SPINACH \cite{SPINACH}, Julia (Quantumoptics.j) \cite{Julia}, and RydIQule \cite{RydIQule}.

After finding the steady state, it is often desirable to study the dependence of the steady state on some parameter $v$, which appears in the Hamiltonian or dissipator. This includes direct dependence on, rate of change with, or an ensemble average over $v$, e.g., Doppler broadening in atomic systems \cite{practicalEIT} or static disorder in condensed matter systems \cite{Trajectory_static_disorder_2010}. In the absence of analytic solutions, it is often required to sample over $v$ and find the steady state for every distinct value. This process is computationally expensive in time and memory. Furthermore, analytic operations, e.g., differentiation and integration with respect to $v$, must be approximated by the corresponding discretized ones, e.g., finite differences and Riemann sums. Reducing the error in these operations necessitates a larger sample size in $v$, so there is a trade-off between the error and computational memory and speed. This presents a computational bottleneck for large open quantum systems.

In the formalism of quantum master equations, the system's Liouville superoperator $\mathcal{L}$ (also known as the Lindbladian) contains all the information about the Hamiltonian and dissipation \cite{QuantumMasterPro}. Consider the common case where the Lindbladian can be decomposed into two parts $\mathcal{L}=\mathcal{L}_0+v\mathcal{L}_1$, for some parameter $v$. This decomposition arises naturally in perturbation theory \cite{dissipation_projected_dynamics_2015,perturb_theory_steady_2011, perturb_theory_steady_2013,Counting_stat_2008} and linear response theory, where $\mathcal{L}_0$ is the original system, and $v\mathcal{L}_1$ is some perturbation. Perturbation theory on open quantum systems has been applied previously in various settings with success, e.g., many-body systems \cite{Perturb_fail_2015, Many_body_perturb_2023}, quantum transport \cite{Perturbation_theory_2020}, and laser cooling of trapped ions \cite{Laser_cooling_perturb_1992}. More generally, systematic perturbative expansions have been developed for open quantum systems with small dissipation \cite{perturb_dissipation_2000,QME_perturbation_2020, perturb_theory_steady_2011}. When combined with adiabatic elimination, perturbation theory can create a simplified effective Liouville superoperator, which only involves the ground states' degrees of freedom \cite{Effective_L_2012, Effective_L2_2012, Sorensen_cavity_2011}. Perturbation theory on open quantum systems has also been used to derive a generalized Kubo formula, which captures the response theory of observables under a perturbation \cite{steady_state_theory_2016,steady_state_theory_2018,response_theory_2021}. 

A general perturbation theory for open quantum systems was introduced by Li \textit{et al}. \cite{perturb_theory_2014}. Given the steady state $\bm{\rho}_0$ of $\mathcal{L}_0$, the theory calculates the perturbed steady state $\bm{\rho}_v$ of $\mathcal{L}_0+v\mathcal{L}_1$ up to any desired power in $v$. In an extension of their work, the authors introduced a partial resummation scheme of the infinite perturbative series for the steady state, where each term in the sum contains corrections up to the infinite order \cite{Perturbation_infinite_resum_2016}. While this work represents a significant improvement over finite-order perturbation theory, it is still perturbative and approximate because the formal infinite series solution has to be truncated for any practical calculation.  In addition to being inherently approximate, perturbation theory can fail even for small perturbations \cite{Perturb_fail_2015} and determining the domain of convergence is a non-trivial problem \cite{perturb_theory_2014, perturb_Gibbs_2018}.

Going beyond perturbation theory, we \rsub{use a generalized inverse of $\mathcal{L}_0$, called the ``Drazin inverse'',} to generate an exact non-perturbative solution for the steady state $\bm{\rho}_v$, for an arbitrary time-independent or periodic perturbation.  We show that by diagonalizing the product of \rsub{the Drazin inverse} of ${\cal L}_0$ and ${\cal L}_1$, the exact $v$-dependence of $\bm{\rho}_v$ can be efficiently obtained.  This method is applicable to a wide variety of open quantum systems.  We give three examples in this paper.

In Sec. \ref{sec:problem_setup}, a summary of the problem is given, as well as a sketch of the main result. In Sec. \ref{sec:propagator_derive}, the main result is derived, which is inspired by Green's function: given the $v=0$ solution $\bm{\rho}_0$, we derive the ``propagator" that acts on the ``zero-case" state to generate the general $v$-dependent solution $\bm{\rho}_v$. The exact non-perturbative solution for all $v$ can be found just by \rsub{a single matrix diagonalization of $\mathcal{L}_0^-\mathcal{L}_1$}, where $\mathcal{L}_0^-$ is a trace-preserving generalized inverse of $\mathcal{L}_0$ \rsub{called the Drazin inverse}. Our derivation is general and does not assume a perturbative series solution. Furthermore, this approach allows for exact analytic operations, such as differentiation and ensemble averaging on $\bm{\rho}_v$ with respect to $v$, without the need for discretization or sampling. The reduction of the number of diagonalizations to \rsub{one} leads to savings in computational time and memory, while still being exact. This approach works even if the system can only be solved numerically. To showcase its validity and utility, we apply the present method to three non-trivial open quantum systems. In Sec. \ref{sec:cavity_opto}, we compute the dependence of cavity optomechanical cooling on the pump laser frequency, just by doing \rsub{a single} diagonalization. In Sec. \ref{sec:magnetometry}, we compute the non-linear response of an atomic magnetometer. In Sec. \ref{sec:Rydberg_sensor}, we compute the velocity dependence and calculate Doppler broadening of a modulated multi-level Rydberg sensor at room temperature, without any sampling. We find agreement between the present method and the exact and numerical solutions for all these systems. \rsub{We also numerically show that the present method achieves a speedup of one to several orders of magnitude, compared to previous approaches, for problems that require large sampling}. This approach works for the steady-state solutions of systems with no or periodic time-dependence, and hence it is expected to be of wide applicability. Sec. \ref{sec:discuss} concludes with a discussion.

All the numerical codes that were used to generate the figures in this work and implement the method are publicly available in Ref. \cite{nagib_exact_2025}.

\section{Problem setup and summary}\label{sec:problem_setup}

We consider a quantum system \rsub{with Hilbert space dimension $d$} whose time evolution is governed by a quantum master equation. In the superoperator form, it is given by \cite{QuantumMasterPro}:
\begin{equation}\label{QME}
\dot{\bm{\rho} }=\mathcal{L}\bm{\rho}
\end{equation}
where $\bm{\rho}$ is the ``vectorized'' version of the density matrix $\rho$, and $\mathcal{L}$ is the Liouville superoperator. A vectorization ${\rm vec}(\rho)$ maps the \rsub{$d \times d$} density matrix $\rho=\sum_{ij} \rho_{ij} \ket{i} \otimes \bra{j}$ into the \rsub{$d^2 \times 1$} column vector $\bm{\rho}={\rm vec}(\rho)= \sum_{ij} \rho_{ij} \ket{i}\otimes\ket{j}$. $\mathcal{L}$ is constructed from the Hamiltonian $H$ and the Lindblad operators $L_k$ that couple the system to the environment, causing dissipation and decoherence. \rsub{Since $\mathcal{L}$ is a superoperator that maps a density matrix into another, its dimension $N$ is quadratic in the Hilbert space, i.e., $\dim(\mathcal{L})=N \times N$ with $N=d^2$}. See Appendix \ref{appendix:Liouville} for an overview of the Liouville superoperator formalism as used in this work. The particular form of $\mathcal{L}$ depends on the nature of the coupling between the system and the environment. For Markovian dynamics, it is given by the Lindblad form: 
\begin{align}\label{louie}
&\mathcal{L}= -i\big(H \otimes\mathbb{1} -\mathbb{1}\otimes H^{\rm T} \big) \\ & \nonumber + \sum_k  \big(L_k \otimes L_k^* - \dfrac{1}{2}\big[L_k^{\dagger} L_k\otimes  \mathbb{1}+ \mathbb{1}  \otimes L_k^{\rm T} L_k^* \big]\big)
\end{align}
At steady state, Eq. \eqref{QME} becomes $\mathcal{L}\bm{\rho}=0$, and finding the steady state reduces to finding the nullspace vector of $\mathcal{L}$. Now consider a quantity $v$ in the Hamiltonian or the Lindblad operators, such that $\mathcal{L}$ is divided into two parts:
\begin{equation}\label{v_problem}
\mathcal{L} \bm{\rho}_v=(\mathcal{L}_0+v \mathcal{L}_1) \bm{\rho}_v=0    
\end{equation}
where both $\mathcal{L}_0$ and $\mathcal{L}_1$ do not depend on $v$. Suppose that a unique steady-state solution exists for the zero-case problem, i.e., $\mathcal{L}_0 \bm{\rho}_0=0$, which is available either analytically or numerically. The question is how to find the $v$-dependent solution $\bm{\rho}_v$ in terms of $\bm{\rho}_0$. \rsub{Moreover, another goal} is to find a procedure to do analytic operations exactly with respect to $v$ even if $\bm{\rho}_v$ is only available numerically, e.g., find the ensemble average of $\bm{\rho}_v$ over a probability distribution, i.e., $\bar{\bm{\rho}}= \int P(v) \bm{\rho}_v dv$. 

\begin{figure}
   \centering
    \includegraphics[width=0.45\textwidth]{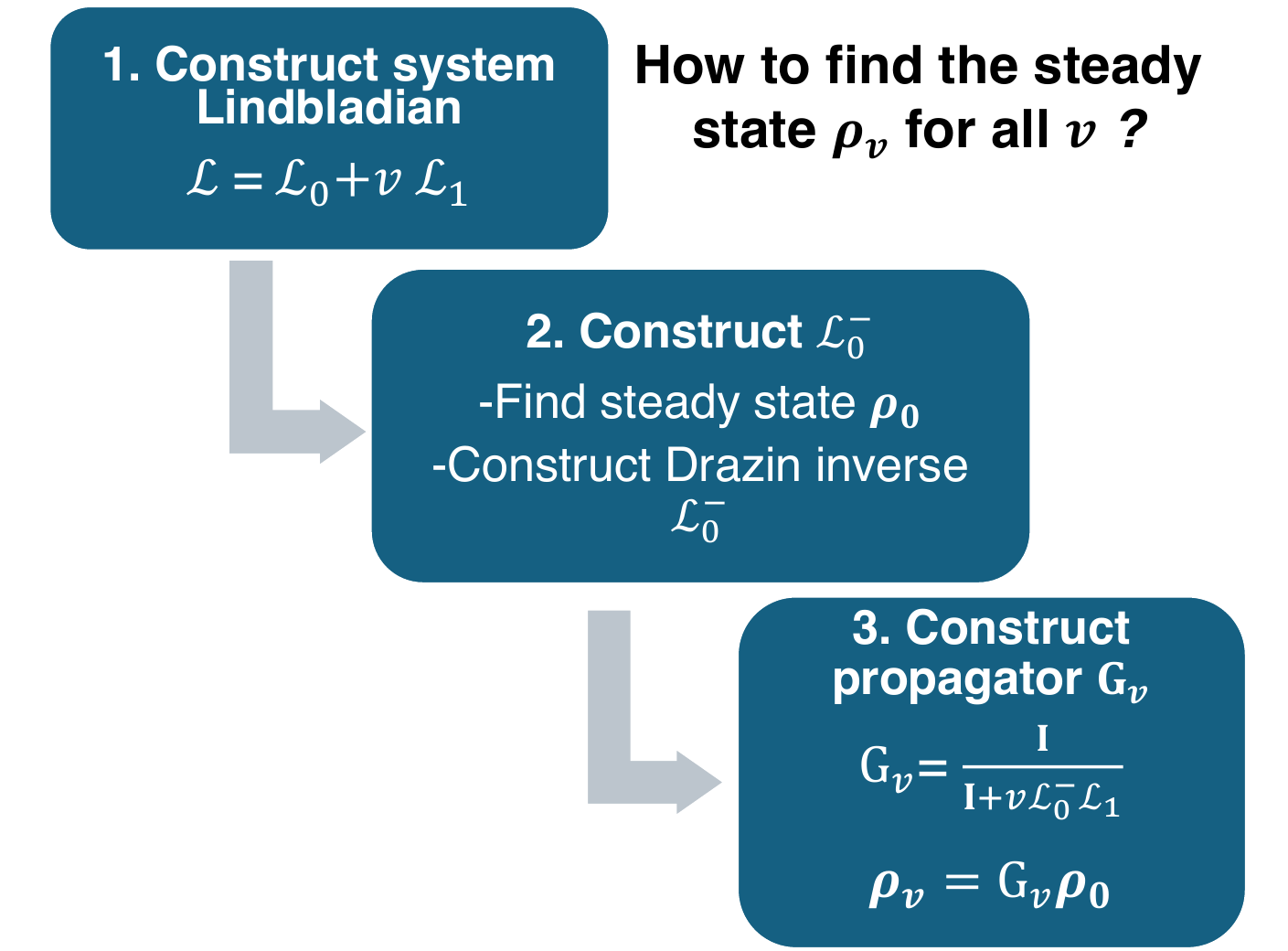}
   \caption{The procedure to find the steady state $\bm{\rho}_{v}$ of an open quantum system, as a function of any system parameter $v$. Once the propagator is constructed, $\bm{\rho}_v$ is \rsub{efficiently} found for all $v$ as well as any analytic operation (e.g., integration and differentiation) on $\bm{\rho}_v$ can be done exactly.}
    \label{fig:idea_diagram}
\end{figure}

Fig. \ref{fig:idea_diagram} pictorially summarizes the main result. We efficiently construct a propagator $G_v$ that acts on the zero-case solution $\bm{\rho}_0$ to generate the $v$-dependent general solution for all $v$, i.e., $G_v \bm{\rho}_0=\bm{\rho}_v$. Once $G_v$ is constructed, not only $\bm{\rho}_v$ is found exactly, but analytic operations, such as differentiation and integration, can be done exactly on $\bm{\rho}_v$, without any sampling. In the next section, we derive $G_v$ and show how to construct it. 

\section{Propagator approach to the nullspace of the Master equation}\label{sec:propagator_derive}

In this section, we show that the propagator $G_v$ that generates the $v$-dependent nullspace  is\begin{equation}\label{Gv}
G_v=(\mathbb{1}+v\mathcal{L}^-_0 \mathcal{L}_1)^{-1} \equiv \dfrac{\mathbb{1}}{\mathbb{1}+v\mathcal{L}^-_0 \mathcal{L}_1}
\end{equation}
where $\mathcal{L}^-_0$ is \rsub{the Drazin} inverse of $\mathcal{L}_0$, to be defined shortly. Let $\bm{R}_g$ and $\bm{L}_g$ be the right and left column eigenvectors of $\mathcal{L}_0$, and $g$ be the corresponding complex eigenvalues. 
The right and left eigenvectors of $\mathcal{L}_0$ obey the following eigenvalue, orthonormality, and completeness relations
\bse\bea
&\mathcal{L}_0 \bm{R}_g=g \bm{R}_g\\
&\bm{L}^{\rm T}_g\mathcal{L}_0 =g\bm{L}^{\rm T}_g\\
&\bm{L}^{\rm T}_i \bm{R}_j=\delta_{ij}\\
&\sum_g  \bm{R}_g  \bm{L}^{\rm T}_g =\mathbb{1}
\eea\ese
In other words, we consider quantum systems that admit an eigendecomposition $\mathcal{L}_0=\sum_{g} g \bm{R}_g \bm{L}_g^T$, where the matrix multiplication $\bm{R}_g \bm{L}_g^T$ is mathematically equivalent to $\bm{R}_g \otimes \bm{L}_g^T$. We shall assume that a unique steady state exists, which is given by the right nullspace vector of $\mathcal{L}_0$, i.e., $\bm{R}_0=\bm{\rho}_0$. The left nullspace vector, $\bm{L}_0$, is the same for all $\mathcal{L}_0$ in quantum systems to ensure probability conservation with time. More precisely, the left nullspace vector is the vectorization of the identity, i.e., $\bm{L}_0={\rm vec}(\mathbb{1})$ for all quantum systems (see Appendix \ref{appendix:Liouville}) \cite{optically_pumped_atoms}.

\rsub{Given that $\mathcal{L}_0$ has a unique nullspace, it is singular and does not admit a matrix inverse in the usual sense. Nonetheless, it is possible to introduce a generalized inverse, the Drazin inverse, which will play a crucial role in solving the perturbation problem [Eq. \eqref{v_problem}] exactly for all perturbations $\mathcal{L}_1$ and all $v$}. $\mathcal{L}^-_0$ is defined as the inverse of $\mathcal{L}_0$ excluding the zero-eigenvector (nullspace):
\begin{equation}\label{L0_minus}
\mathcal{L}^-_0=\sum_{g\neq 0}\dfrac{1}{g}\bm{R}_g  \bm{L}_g^T
\end{equation}
\rsub{An equivalent form that avoids explicit diagonalization is
\begin{equation}\label{L0minus_fast}
\mathcal{L}^-_0=(\mathcal{L}_0+\bm{R}_0  \bm{L}_0^T)^{-1}-\bm{R}_0  \bm{L}_0^T
\end{equation}
which can be numerically more efficient than the construction based on Eq. \eqref{L0_minus} if $\bm{R}_0$ is available or can be obtained efficiently. See Appendix \ref{appnedix:efficient_L0_minus} for more details on the Drazin inverse and its construction.} 

$\mathcal{L}_0$ and $\mathcal{L}^-_0$ obey the \rsub{following key} generalized inverse relations
\begin{equation}\label{generalized_property}
\mathcal{L}_0\mathcal{L}^-_0=\mathcal{L}^-_0\mathcal{L}_0=\mathbb{1}-\bm{R}_0  \bm{L}_0^T
\end{equation}
\rsub{i.e., $\mathcal{L}_0\mathcal{L}^-_0$ is the projection operator to the space orthogonal to the nullspace of $\mathcal{L}_0$}.

Next, we prove that this propagator generates $\bm{\rho}_v$. Acting on Eq. \eqref{v_problem} with $\mathcal{L}^-_0$ and using Eq. \eqref{generalized_property} gives
\begin{equation}
\mathcal{L}^-_0(\mathcal{L}_0+v \mathcal{L}_1) \bm{\rho}_v=(\mathbb{1}-\bm{R}_0  \bm{L}_0^T+v\mathcal{L}^-_0\mathcal{L}_1)\bm{\rho}_v=0
\end{equation}
Using the identity $\bm{L}_0^T \bm{\rho}_v={\rm vec}(\mathbb{1})^{T}\cdot\bm{\rho}_v={\rm Tr}(\rho_v)=1$, which is a statement of probability conservation (see Appendix \ref{appendix:Liouville} for proof) \cite{optically_pumped_atoms}, setting $\bm{R}_0=\bm{\rho}_0$, and rearranging, we get:
\begin{equation}
(\mathbb{1}+v\mathcal{L}^-_0\mathcal{L}_{1})\bm{\rho}_v=\bm{\rho}_0 
\end{equation}
Inverting this equation readily gives:
\begin{equation}\label{rho_v}
\bm{\rho}_v=\dfrac{\mathbb{1}}{\mathbb{1}+v\mathcal{L}^-_0 \mathcal{L}_1}\bm{\rho}_0
\end{equation}
which shows that Eq. \eqref{Gv} indeed gives the correct propagator. This equation can be generalized for the case of more than one variable $\{v_i\}$, as described in Appendix \ref{appnedix:L0_minus_generalized}. 

To be able to deal analytically with Eq. \eqref{rho_v}, we diagonalize $\mathcal{L}^-_0 \mathcal{L}_1$ to find its right and left eigenvectors, $\bm{r}_\lambda$ and $\bm{l}_\lambda$, and their corresponding eigenvalues $\lambda$. \rsub{We shall denote the right and left eigenvectors matrices of $\mathcal{L}_0^-\mathcal{L}_1$ as $r=(\bm{r}_1,...,\bm{r}_N)$ and $l=(\bm{l}_1,...,\bm{l}_N)$, and the corresponding eigenvalues as the vector $\Lambda=(\lambda_1,...,\lambda_N)$. The left eigenvectors can be computed as $l=(r^{\rm T})^{-1}$ or $l^{\rm T}=r^{-1}$.} Then the eigendecomposition of the propagator \rsub{and the steady state} in that basis gives:
\bse\bea 
\label{Gv_eigen}G_v&=&\sum_{\lambda=\lambda_1}^{\lambda_N} \dfrac{1}{1+\lambda v} \bm{r}_\lambda \bm{l}^{\rm T}_\lambda=r \ {\rm diag}\left(\dfrac{1}{1+v \Lambda}\right) l^{\rm T} \\ \label{rhov_1D}
\bm{\rho}_v&=& G_v \bm{\rho}_0=r \ {\rm diag}\left(\dfrac{1}{1+v \Lambda}\right) (l^{\rm T}\bm{\rho}_0)
\eea\ese
\rsub{In this new eigenbasis of $\mathcal{L}^-_0 \mathcal{L}_1$, the steady state $\bm{\rho}_v$ is simply parametrized by the eigenvalues, where $\lambda_i$ and $l^{\rm T}\bm{\rho}_0$ are precomputed only once. Computing the steady state for any new $v$ then amounts to merely updating the $N=d^2$ scalar quantities $1/(1+v \lambda_i)$. In Appendix \ref{appnedix:efficient_Gv}, we show that in fact only the nonzero eigenvalues $\lambda_i \neq 0$ contribute to the perturbed steady state, which leads to a further speedup.}

\rsub{To transform a vectorized operator from the original basis to the new basis, we do the mapping ${\rm vec}(A) \rightarrow  l^{\rm T}{\rm vec}(A)$ and ${\rm vec}(A)^{\dagger} \rightarrow  {\rm vec}(A)^{\dagger} \ r$ for the dual space. For example, computing the expectation value of a Hermitian operator $A$ with respect to the state $\bm{\rho}_v$ becomes:}
\rsub{
\begin{equation}\label{A_avg}
\braket{A}=\left({\rm vec}(A)^{\dagger} \ r\right){\rm diag}\left(\dfrac{1}{1+v \Lambda}\right) (l^{\rm T}\bm{\rho}_0)
\end{equation}}
\rsub{where the vectorized operators ${\rm vec}(A)$ and $\bm{\rho}_0$ are expressed in the new eigenbasis and precomputed once for all $v$. Computing the expectation value for any $v$ is achieved by simply updating the $N$ scalar terms in the equation above and carrying out the matrix-vector multiplication.}

\rsub{Since $\mathcal{L}_0^-$} is trace-preserving, $\bm{\rho}_v$ is automatically normalized if $\bm{\rho}_0$ is normalized. We have obtained an exact non-perturbative solution for an arbitrary perturbation $v \mathcal{L}_1$, generated from the unperturbed steady state. The problem of finding the arbitrary steady state for all $v$ has thus been reduced to \rsub{constructing $\mathcal{L}_0^-$ and a single} diagonalization of $\mathcal{L}^-_0 \mathcal{L}_1$. This diagonalization, in turn, can be done either numerically or analytically.  Once the diagonalization is done, $\bm{\rho}_v$ follows immediately from Eqs. \eqref{Gv_eigen} and \eqref{rhov_1D}, without the need to redo the diagonalization for every distinct $v$.

It is important to emphasize the generality of this derivation: no assumptions were made about the magnitude of $v$ or whether $\bm{\rho}_v$ admits a perturbative expansion in $v$. In Appendix \ref{appendix:perturbation_theory}, we elaborate on the relation between this method and perturbation theory on open quantum systems. \rsub{While a perturbative series always has a finite radius of convergence in $v$, i.e., it converges only when $|v \lambda_i|<1$ for all $\lambda_i$, the present result generates the exact solution for all $v$.} \rsub{Moreover}, for cases where a perturbative expansion exists, the \rsub{present} solution is equivalent to perturbation theory with corrections up to the infinite order. Therefore, this solution goes well beyond perturbation theory \cite{perturb_theory_2014, Perturbation_infinite_resum_2016}.

In addition to being exact, the present method can offer computational speedup in many cases. \rsub{For exact (e.g., LU decomposition and Arnoldi/Krylov iterations) or approximate (e.g., perturbation theory or variational perturbation theory) methods, the computational cost to find the steady state for $P$ samples of a generic $N \times N$ Lindbladian scales as $O(PN^3)$ or $O(Pd^6)$ \cite{VPT2025}. Under the present method, on the other hand, there is a constant overhead associated with \rsub{constructing the Drazin inverse and diagonalizing $\mathcal{L}_0^-\mathcal{L}_1$}, independent of the number of samples in $v$, which scales as $O(N^3)=O(d^6)$. After this step, any subsequent $\bm{\rho}_{v}$ can be generated efficiently by updating the $N$ scalar terms in Eq. \eqref{rhov_1D}, i.e., the cost is $O(PN)$ for $P$ samples. Thus, the total cost of the present approach is estimated to be $O(N^3+PN)=O(d^6+Pd^2)$ compared to $O(Pd^6)$ in previous approaches. In the next sections, we will give numerical examples where the present method achieves a significant speedup for problems that require large samples.} 

Since $v$ appears in a simple form in Eq. \eqref{Gv_eigen}, analytic operations with respect to $v$ on $\bm{\rho}_v$ can be done exactly, e.g., differentiation and integration with respect to $v$. These operations can be done exactly, even when only a numerical solution is available, without the need to approximate these operations using finite differences and Riemann sums. For example, an ensemble average can be analytically computed by integrating the $v$-dependence in $G_v$, i.e., $\int dv P(v) 1/(1+\lambda v)$. We obtain the ensemble average exactly without any approximations, while simultaneously saving significant computational resources by avoiding sampling over $v$. For the case where $P(v)$ is a Gaussian distribution with a standard deviation $\sigma_v$, i.e., $P(v)=(2\pi \sigma_v^2)^{-1/2}\exp(-v^2/ 2\sigma_v^2)$, the $v$-dependence is integrated, and we get the ensemble-averaged steady state
\begin{align}\label{rho_gauss_1D}
&\bar{\bm{\rho}}= \sum_{\lambda=0} \bm{r}_\lambda  (\bm{l}^{\rm T}_\lambda \bm{\rho}_0) \\ &\nonumber+ \sum_{\lambda\neq 0} \dfrac{\sqrt{\pi/2}}{\sqrt{-\lambda^2} \sigma_v}   e^{-\frac{1}{2 \lambda^2 \sigma_v^2}}\bigg(1+{\rm erf}\bigg[\dfrac{\sqrt{-\lambda^2}}{\sqrt{2} \lambda^2 \sigma_v }\bigg]\bigg)   \bm{r}_\lambda  (\bm{l}^{\rm T}_\lambda \bm{\rho}_0)
\end{align}
where $\rm erf$ is the error function, and the first and the second sums are over the zero and non-zero eigenvalues $\lambda$, respectively. In Appendix \ref{appendix:doppler}, we calculate Doppler broadening for a two-level system, for which an exact solution is analytically available, showing agreement between the present and the analytic solutions. In Appendix \ref{appendix:avg}, we explicitly compute the ensemble average for a Lorentzian distribution. This approach can also be applied to compute the exact ensemble average with respect to any other desired distribution.

In the next sections, we apply this approach to three non-trivial open quantum systems, showing that it agrees with the exact results, as well as achieving computational speedup for calculations that require large sampling. 

\begin{figure}
   \centering
    \includegraphics[width=0.45\textwidth]{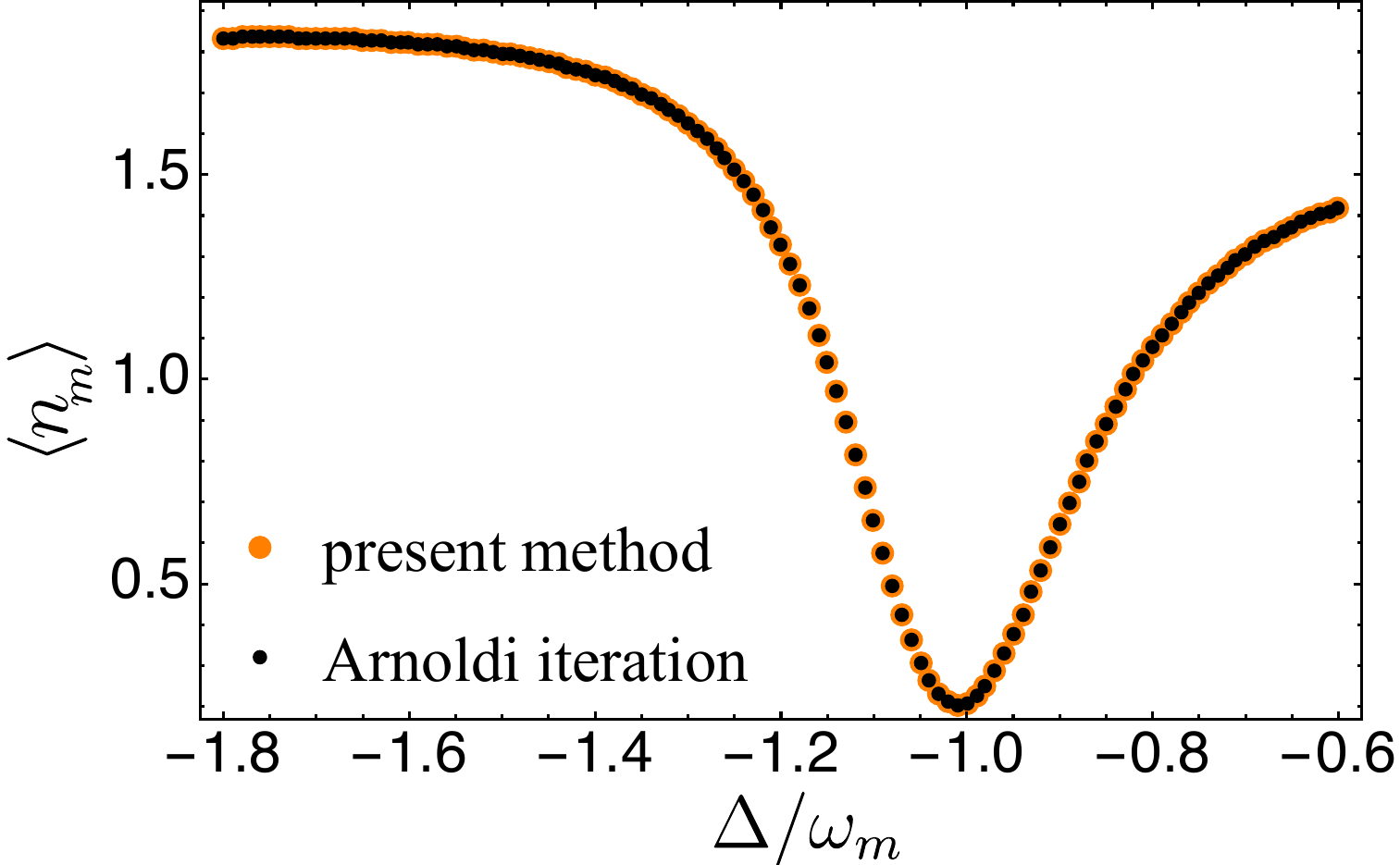}
   \caption{Phonon number $\braket{n_m}$ at steady state versus detuning $\Delta$, using the present method (orange) and \rsub{the sparse Arnoldi iteration} (black). The system parameters used are (arbitrary units): $\omega_m=10,\ g=1, \ \eta=2, \ \kappa=1, \ \Gamma_m=0.015,$ and $n_{\rm th}=2$, where the parameters choice was inspired from Ref. \cite{qojulia_optomech_cooling}. Fock space dimensions of $N_m=10$ and $N_c=4$ were used for the simulation of the oscillator and the cavity, respectively.}
    \label{fig:Nm}
\end{figure}
\section{Example 1: Cavity optomechanical cooling}\label{sec:cavity_opto}

The standard Hamiltonian coupling a laser-driven optical cavity to a mechanical oscillator, in a frame rotating with the laser frequency, is given by \cite{Cavityoptomechanics,opto_cool}
\begin{equation}\label{H_cavity_optp}
H= -\Delta a^{\dagger}a+\omega_m b^{\dagger}b-g_0 a^{\dagger}a(b+b^{\dagger})+ \eta(a+a^{\dagger})
\end{equation}
where $a$ ($b$) is the annihilation operator for the cavity (mechanical oscillator), and $g_0$ is the vacuum coupling strength between the cavity and the oscillator. $\omega_c$ ($\omega_m$) is the cavity (oscillator) resonance frequency, $\Delta=\omega_L-\omega_c$ is the detuning between the driving laser and the cavity, and $\eta$ is the laser pump strength. There is a 
Lindblad operator associated with the cavity photon losses $L_{c}$, as well as operators $L^{\pm}_{m}$ that couple the oscillator to a thermal bath:
\begin{subequations}\label{lindbald_cavity}
\begin{align}
&L_{c}=\sqrt{\kappa}a\\
&L^{+}_{m}=\sqrt{\Gamma_m n_{\rm th}}b^{\dagger}\\
&L^{-}_{m}=\sqrt{\Gamma_m (n_{\rm th}+1)}b
\end{align}
\end{subequations}
where $\kappa$ is the cavity decay rate, $\Gamma_m$ is the mechanical damping rate, and $n_{\rm th}$ is the average thermal phonon number. When $\Delta=-\omega_m$ and $\omega_m \gg \kappa$, the cavity and the oscillator have the same resonance frequencies and can interchange photons and phonons. As the phonons get transferred as photons into the cavity, they are subsequently expelled outside the system. This mechanism leads to the cooling of the oscillator. Increasing the pump strength $\eta$ enhances the effective interaction strength between the two resonators, driving the steady-state phonon number down \cite{Cavityoptomechanics, opto_cool}.

\begin{figure}
   \centering
    \includegraphics[width=0.47\textwidth]{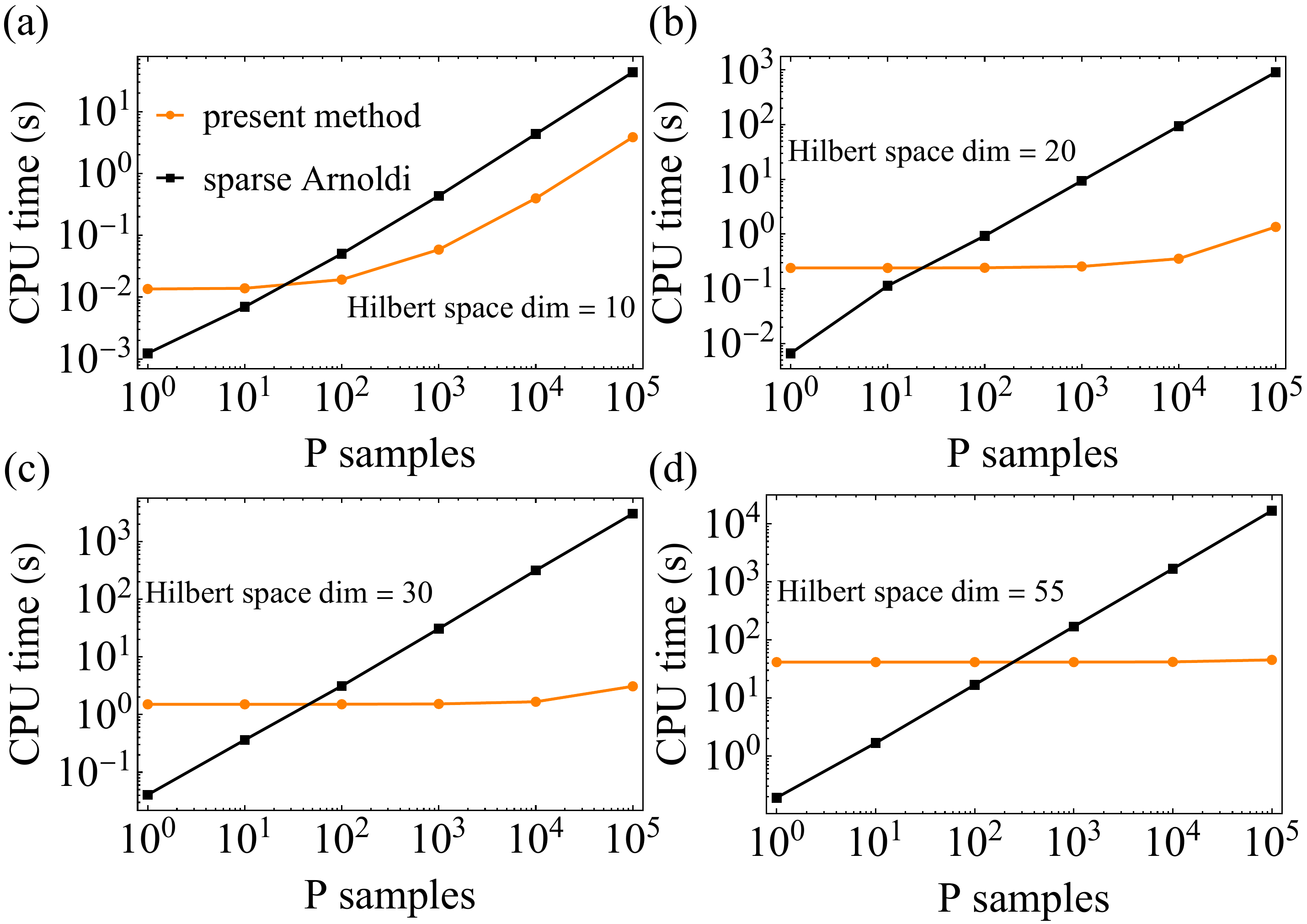}
   \caption{Solution time to find the steady state of the cavity optomechanical system versus the number of samples $P$, using the present method (orange) and the sparse Arnoldi iteration (black), for different Hilbert space dimensions.}
    \label{fig:PCPUtime}
\end{figure}

\rsub{To benchmark the present method in terms of accuracy and speed, we will compare it with the sparse Arnoldi iteration \cite{Review_2015, QuantumMasterPro}, which is a standard method to find the nullspace of open quantum systems. Since it is an iterative method that can exploit the sparse structure of a Lindbladian, avoiding a full diagonalization, it is numerically efficient \footnote{In MATHEMATICA, one can find the nullspace of a Lindbladian \texttt{L} by calling \texttt{Eigenvectors[L, 1, Method -> {"Arnoldi", "Shift" -> 0}]}, and embedding \texttt{L} in the data structure \texttt{SparseArray} to exploit sparsity if present.}.}

First, we apply the present method to compute the dependence of the steady-state average phonon number $\braket{n_m}$ on the laser detuning $\Delta$. $\mathcal{L}_0$ is constructed from $H$ and the Lindblad operators [Eqs. \eqref{H_cavity_optp} and \eqref{lindbald_cavity}] with $\Delta=-\omega_m$, using Eq. \eqref{louie}. \rsub{The the steady state at $\Delta=-\omega_m$, i.e., $\bm{\rho}_0$, is numerically found (using any method such as sparse Arnoldi)}. Then the \rsub{Drazin} inverse $\mathcal{L}_0^-$ is constructed [Eq. \eqref{L0minus_fast}]. To find the steady state $\bm{\rho}_{\delta}$ at any other detuning $\delta$ away from $\omega_m$, i.e, $\Delta=-\omega_m+\delta$, we note that the corresponding part in the Hamiltonian, $-\Delta a^{\dag} a$, can be decomposed as $\omega_m a^{\dag} a-\delta a^{\dag}a$. Thus, we make the identification $H_1=- a^{\dag}a$ and construct $\mathcal{L}_1$. Finally, we numerically diagonalize $\mathcal{L}_0^{-}\mathcal{L}_1$ to find the propagator $G_{\delta}$ [Eq. \eqref{Gv_eigen}], which generates the $\delta$-dependent exact solution [Eq \eqref{rhov_1D}]:
\begin{equation}\label{rho_delta}
\bm{\rho}_{\delta}=\sum_\lambda \dfrac{1}{1+\lambda \delta} \bm{r}_\lambda (\bm{l}^{\rm T}_\lambda\bm{\rho}_{0})
\end{equation}
where $\bm{\rho}_0$ is the solution at $\delta=0$ (or equivalently $\Delta=-\omega_m$). \rsub{In the actual numerical simulation, we work in the eigenbasis of $\mathcal{L}_0^- \mathcal{L}_1$ [Eq. \eqref{rhov_1D}].} Fig. \ref{fig:Nm} shows the steady state phonon number $\braket{n_m}$ versus the detuning $\Delta$, computed using the present method (orange) and \rsub{the sparse Arnoldi iteration (black)}. In the \rsub{sparse Arnoldi} approach, a new $\mathcal{L}$ is initialized for every $\Delta$, and the steady state is found by numerically computing the nullspace for that $\mathcal{L}$. Using the present method, on the other hand, only \rsub{a single} diagonalization, that of $\mathcal{L}_0^{-}\mathcal{L}_1$, was used to generate the entire solution for the 121 different values of $\Delta$ plotted [Eq. \eqref{rho_delta}]. Both methods are in excellent agreement with each other.

\rsub{Fig. \ref{fig:PCPUtime} compares the solution time using the present method (orange) and the sparse Arnoldi iteration (black), as a function of the number of samples $P$. Each subfigure shows the solution time for a different dimension of the Hilbert space $d$. In the sparse Arnoldi approach, the solution time increases linearly with $P$, as expected. Moreover, for a given $P$ the solution time increases with $d$. In the present approach the solution time is almost independent of $P$, except for small $d\le 10$. The cost is dominated by the constant overhead in constructing $\mathcal{L}_0^-$ and diagonalizing $\mathcal{L}_0^-\mathcal{L}_1$ while the cost to generate $P$ solutions is relatively negligible. The present approach gains a relative speedup even for small sample sizes between $P \sim10^1-10^2$, and the achieved speedup is between one and three orders of magnitudes for larger samples.}

\section{Example 2: Magnetometry} \label{sec:magnetometry}

\begin{figure}
   \centering
    \includegraphics[width=0.45\textwidth]{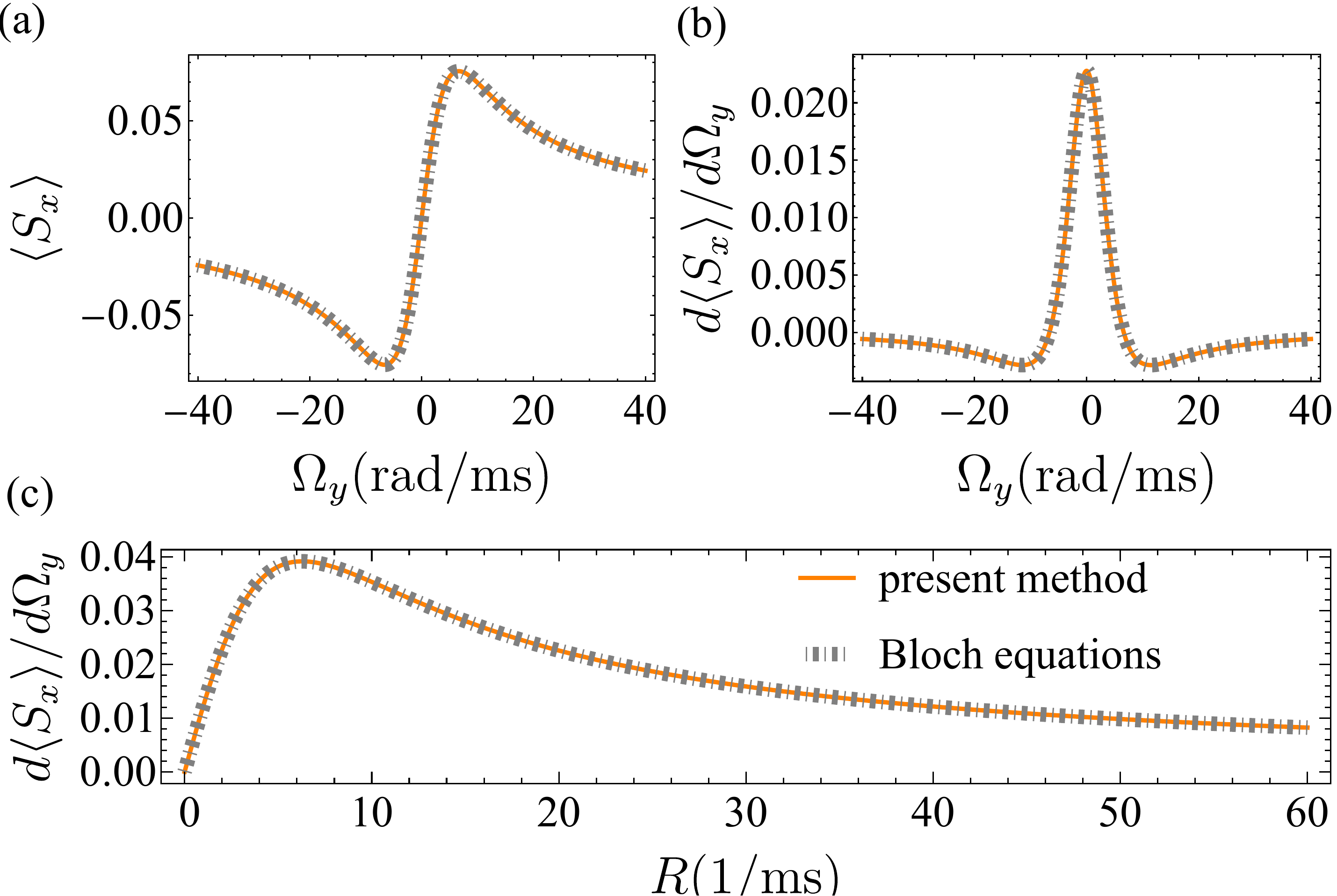}
   \caption{(a) Magnetometer response $\braket{S_x}$ as a function of $\Omega_y$ for $\Omega_z=2\pi\times1$ kHz. The optical pumping and randomization rates are $R=2000/$s and $\Gamma=100/$s, respectively. Dashed lines show analytic Bloch equation solutions. (b) Rate of change of $\braket{S_x}$ with $\Omega_y$. The entire response in (a) and (b) has been calculated \rsub{using} [Eq. \eqref{propmag}]. (c) The magnetometer sensitivity versus $R$, computed analytically by Eq. \eqref{magnetometer_optimize}.}
    \label{fig:magfig}
\end{figure}

We next consider a different problem, the magnetic response of $^{87}$Rb atoms in an optically pumped buffer gas.  The Hamiltonian is \cite{Thad_optica_pump_review}
\begin{equation}
H={\omega_0 \over I+1/2}{\bf S}\cdot{\bf I}+\gamma_S (S_y B_y+S_z B_z)=H_0+\Omega_y S_y
\end{equation}
where $S=1/2$ is the electron spin, $I=3/2$ is the nuclear spin, and $\omega_0=2\pi \times 6.8 $ GHz is the hyperfine splitting. Atoms are subjected to optical pumping along the $z$-axis at a rate $R$ and electron spin-randomization collisions at a rate $\Gamma$.  The $64\times 64$ Liouville superoperator is ${\cal L}={\cal L}_0+\Omega_y{\cal L}_1$ in the form of Eq. (2) with Lindblad operators
\begin{equation}
\{L_1\ , \dots, L_5\}=\left\{\sqrt{\Gamma} S_x,\sqrt{\Gamma} S_y,\sqrt{\Gamma} S_z,\sqrt{R}S_+,\sqrt{R}S_z\right\}
\end{equation}
This is a model of a low--field magnetometer, including the full hyperfine structure. As in the previous example, we can find the complete non-linear response of the atoms to the transverse magnetic field by  constructing the Drazin inverse ${\cal L}_0^-$, and then numerically diagonalizing ${\cal L}_0^-{\cal L}_1$ to find its eigenvalues $\lambda$ and the corresponding right and left eigenvectors, $\bm{r}_\lambda$ and $\bm{l}_\lambda$.  From these we again construct the propagator
\begin{equation}\label{propmag}
G_\Omega=\sum_\lambda \dfrac{1}{1+\lambda \Omega_y} (\bm{r}_\lambda \bm{l}^{\rm T}_\lambda)
\end{equation}
to enable finding the steady-state density matrix for arbitrary $\Omega_y$ from the nullspace $\bm{\rho}_0$ of ${\cal L}_0$.  We re-emphasize that the calculation is non-perturbative, and with a \rsub{single} matrix diagonalization the steady-state is completely determined for arbitrary values of $\Omega_y$. 
Fig. \ref{fig:magfig} (a) shows the magnetometer response $\braket{S_x}$ as a function of the field $\Omega_y$, superposed on the steady-state solution to the Bloch equations (cf. Appendix \ref{appendix:Bloch} for the analytic solution). There is excellent agreement between the present approach and the analytic solution for all values of $\Omega_y$. As described previously, this method also allows for analytic operations (e.g., differentiation) to be done exactly without the need for numerical discretization or sampling. Fig. \ref{fig:magfig} (b) shows the rate of change of $\braket{S_x}$ versus $\Omega_y$, computed by directly taking the derivative of the propagator [Eq. \eqref{propmag}] with respect to $\Omega_y$. 

In certain cases, this approach allows for analytic optimization, even when only numerical steady-state solutions are available. As an example, consider the magnetometer response given by $d\braket{S_x}/d\Omega_y={\rm vec}(S_x)\cdot\partial\bm{\rho}(R, \Omega_y=0)/\partial\Omega_y$. This measures the magnetometer sensitivity to the field $\Omega_y$ as a function of $R$. To optimize for $R$ by brute force, one needs to find the steady state $\bm{\rho}(R,\Omega_y)$ for a sample of $R$ and $\Omega_y$ (around $\Omega_y=0$), and then approximate the derivative as a finite difference. Alternatively, using this approach, we show how to obtain an exact analytic expression for $\partial_\Omega \bm{\rho}(R, \Omega_y=0)$ as a function of $R$. This way, the magnetometer can directly be optimized for $R$. In Appendix \ref{appenddix:optimize_magnetometer}, we show that the magnetometer sensitivity is given by
\begin{align}\label{magnetometer_optimize}
\nonumber &\dfrac{\partial\bm{\rho}(R, 0)}{\partial \Omega_y} =  -\frac{\mathbb{1}}{\mathbb{1} + R \mathcal{L}_0^- \mathcal{L}_{R}} \mathcal{L}_0^- \mathcal{L}_{\Omega} \bm{\rho}(R, 0) \\ & = -\frac{\mathbb{1}}{\mathbb{1} + R \mathcal{L}_0^- \mathcal{L}_{R}} \mathcal{L}_0^- \mathcal{L}_{\Omega} \frac{\mathbb{1}}{\mathbb{1} + R \mathcal{L}_0^- \mathcal{L}_{R}} \bm{\rho}(0,0)
\end{align}
where $\mathcal{L}_\Omega$ is the superoperator constructed from $S_y$, and $\mathcal{L}_R$ is constructed from the Lindblad operators $S_{+}$ and $S_z$, using Eq. \eqref{louie}. $\bm{\rho}(0,0)$ is the steady state at $\Omega_y=R=0$. Once all the quantities in the equation above are computed, $\partial_\Omega \bm{\rho}(R, \Omega_y=0)$ is found for all $R$. This expression can be efficiently computed when expressed in the eigenbasis of $\mathcal{L}_0^- \mathcal{L}_{R}$ (see Appendix \ref{appnedix:efficient_Gv}). Using Eq. \eqref{magnetometer_optimize}, the maximum sensitivity is found at $R=6.28/{\rm ms}$. Fig. \ref{fig:magfig}(c) shows excellent agreement between this approach [Eq. \eqref{magnetometer_optimize}] and the analytic solution for all $R$. \rsub{One important application of the derivative formula above is in the efficient estimation of an unknown constant parameter $\phi$ in a Lindbladian $\mathcal{L}(\phi,\theta)$, where $\theta$ is a known controllable parameter \cite{VPT2025}}.

\section{Example 3: Doppler broadening of a Rydberg sensor}\label{sec:Rydberg_sensor}

\begin{figure}
   \centering
    \includegraphics[width=0.45\textwidth]{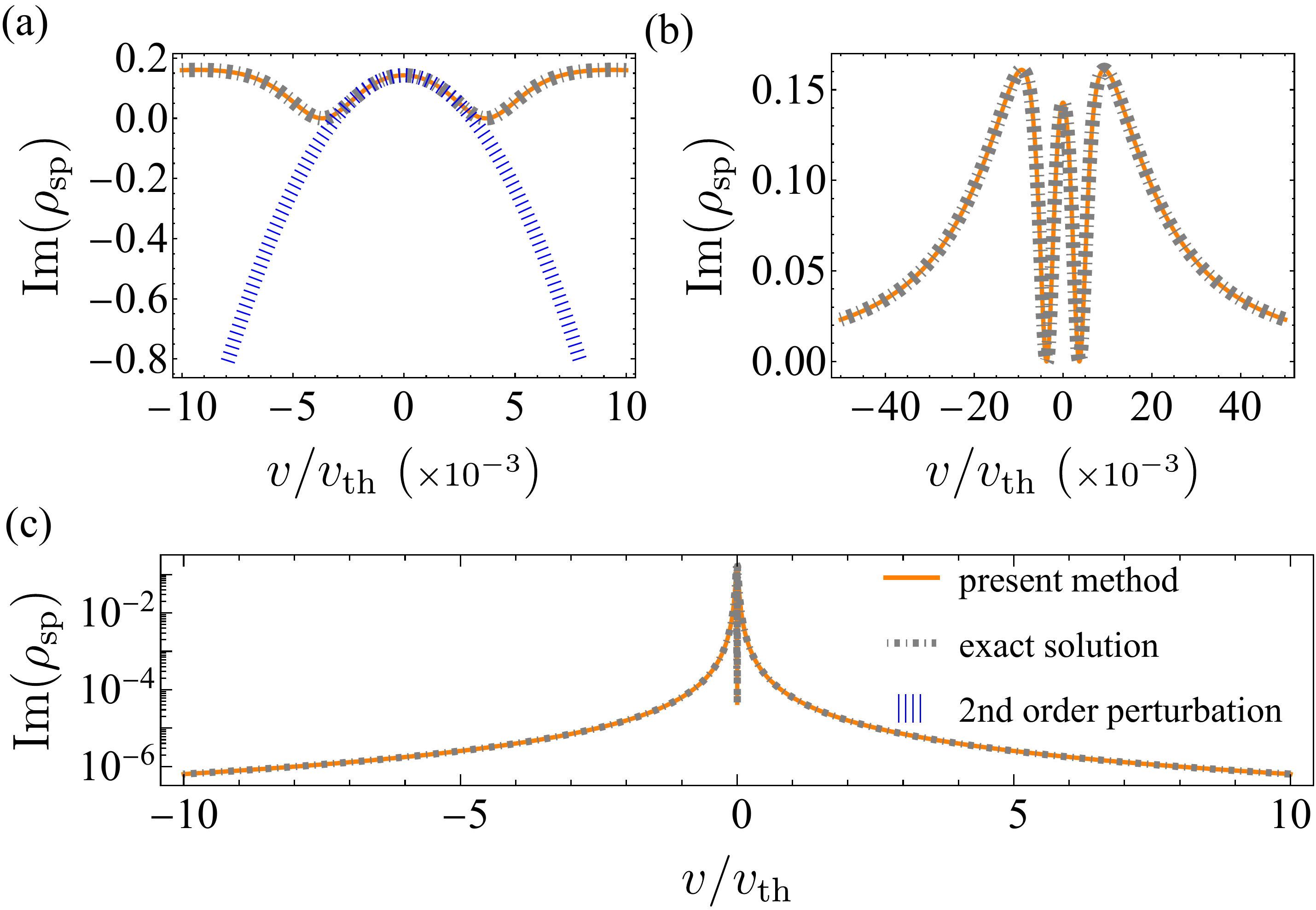}
   \caption{(a) ${\rm Im}(\rho_{\rm sp})$ as a function of velocity using the present method (orange solid), exact solution (dashed gray), and second-order perturbation theory (dotted blue). The system parameters are: thermal velocity $v_{\rm th}=169.5 \ {\rm \mu m/\mu s}$, $\Omega_1=2\pi \times 2 \ {\rm MHz},$ $\Omega_2=\Omega_3=2\pi \times1 \ {\rm MHz}$ and $\Gamma=2\pi \times 6 \ {\rm MHz}$. The modulation frequency is set to $f=0 \ {\rm MHz}$ here. $k_1/2\pi=1/(0.78\rm{\ \mu m})$ and $k_2/2\pi=-1/(1.248 \ {\rm  \mu m})$. The detunings are set in the EIT resonance regime $\Delta=\delta=0$. Panles (b) and (c) show the agreement between the present and the exact solution over a larger domain range of $v$.}
    \label{fig:rho_sp_v}
\end{figure}

For a third example, we consider a multi-level system with time modulation, which experiences electromagnetically induced transparency (EIT) and Autler-Townes (AT) splitting. An ensemble of such atoms placed in a room-temperature vapor cell is used as a sensor to measure electromagnetic fields \cite{practicalEIT}. Since decoherence from Doppler broadening limits the performance of atomic sensors, simulating its effects accurately is an important problem \cite{Raman_Ramsey, practicalEIT}. For sufficiently complicated systems, calculation of Doppler broadening presents a bottleneck in computation time and memory, increasing the simulation time by several orders of magnitude \cite{RydIQule}. It is this problem that originally motivated this work.  We begin by showing that the present method agrees with the exact solution in describing the non-trivial velocity dependence of the system. Then we calculate the Doppler broadening efficiently and accurately, without any sampling. 

\subsection{Present method versus exact solution and perturbation theory}

Consider a four-level ladder system in $\ce{^{87}Rb}$, with a ground state $\rm S$, an excited state $\rm P$, and Rydberg states $\rm D$ and $\rm F$. Let $\Delta$ be the one-photon detuning between $\rm S$ and $\rm P$, $\delta$ the two-photon detuning between $\rm S$ and the Rydberg states. There is a probe field  $\Omega_1$ coupling $\rm S$ and $\rm P$, and $\Omega_2(t)$ is the control field coupling $\rm P$ and $\rm D$. The field $\Omega_3$ \rsub{resonantly} couples the Rydberg states $\rm D $ and $\rm F$. The control field is time-modulated harmonically, i.e., $\Omega_2(t)=\Omega_2 \cos(\omega t)$, where $\Omega_2$ and $\omega= 2 \pi f$ are the modulation amplitude and frequency, respectively. $\rm P$ decays to $\rm S$ at a rate $\Gamma=2\pi \times 6 \ \rm{MHz}$. The Hamiltonian and Lindblad operator for this system read \cite{practicalEIT}:
\begin{align}\label{H_Rydberg}
&H=-\Delta \ \ket{\rm p}\bra{\rm p} -\delta  \ket{\rm d}\bra{\rm d} -\delta  \ket{\rm f}\bra{\rm f}+\dfrac{\Omega_1}{2} \ket{\rm s}\bra{\rm p} \\ &\nonumber +\dfrac{\Omega_2(t)}{2} \ket{\rm p}\bra{\rm d}+\dfrac{\Omega_3}{2} \ket{\rm f}\bra{\rm d} +\rm h.c.\\ & L=\sqrt{\Gamma}\ket{\rm s}\Bra{ \rm p}
\end{align}
from which $\mathcal{L}_0$ can be readily constructed. At non-zero velocity $v \neq 0$, the atom experiences Doppler shift for the one- and two-photon detunings, $\Delta \rightarrow \Delta-k_1 v$ and $\delta \rightarrow \delta-k_2 v$, where $k_1$ and $k_2$ are the effective wavenumbers for the one- and two-photon transitions. To determine the exact velocity dependence of the steady state, we proceed as before. First, we solve the zero-velocity problem by \rsub{finding} $\bm{\rho}_0$ and \rsub{constructing} $\mathcal{L}_0^-$. Next, we diagonalize $\mathcal{L}_0^- \mathcal{L}_1$ to get their eigenvalues and eigenvectors. Here, $\mathcal{L}_1$ is the Doppler-shifted component of $\mathcal{L}$ given by $\mathcal{L}_1=d\mathcal{L}_0(\Delta-k_1 v,\delta-k_2 v )/dv$. Then $\bm{\rho}_v$ would be given by Eqs. \eqref{Gv_eigen} and \eqref{rhov_1D} for all $v$. To be able to assess the exactness of the present method, we restrict this subsection to the case of no modulation (i.e., $f=0$), for which an exact analytic solution can be found. The exact solution can be computed analytically by Mathematica's native Nullspace function for $v=0$, and the $v$-dependence can be generated by applying the shifts $\Delta \rightarrow \Delta-k_1 v$ and $\delta \rightarrow \delta-k_2 v$ to the $v=0$ solution. In atomic sensors, the signal of interest is the imaginary component of the coherence between the ground and excited state, ${\rm Im}(\rho_{\rm sp})$, which is proportional to the absorption coefficient of the probe field. Fig. \ref{fig:rho_sp_v} shows ${\rm Im}(\rho_{\rm sp})$ versus the velocity (normalized by the thermal velocity $v_{\rm th}$), using the present method, exact solution, and second-order perturbation theory (cf. Appendix \ref{appendix:perturbation_theory}). The figure shows excellent agreement between the present approach and the exact solution, which is able to reproduce the exact EIT and AT structure at all scales of $v$. We reemphasize that only \rsub{a single} diagonalization was used to generate the entire $v$-dependence. The present approach works well beyond the domain of perturbation theory, where the latter only agrees with the exact solution in a very small interval of $v/v_{\rm th} \approx  5 \times 10^{-3}$.

\begin{figure}
   \centering\includegraphics[width=\linewidth]{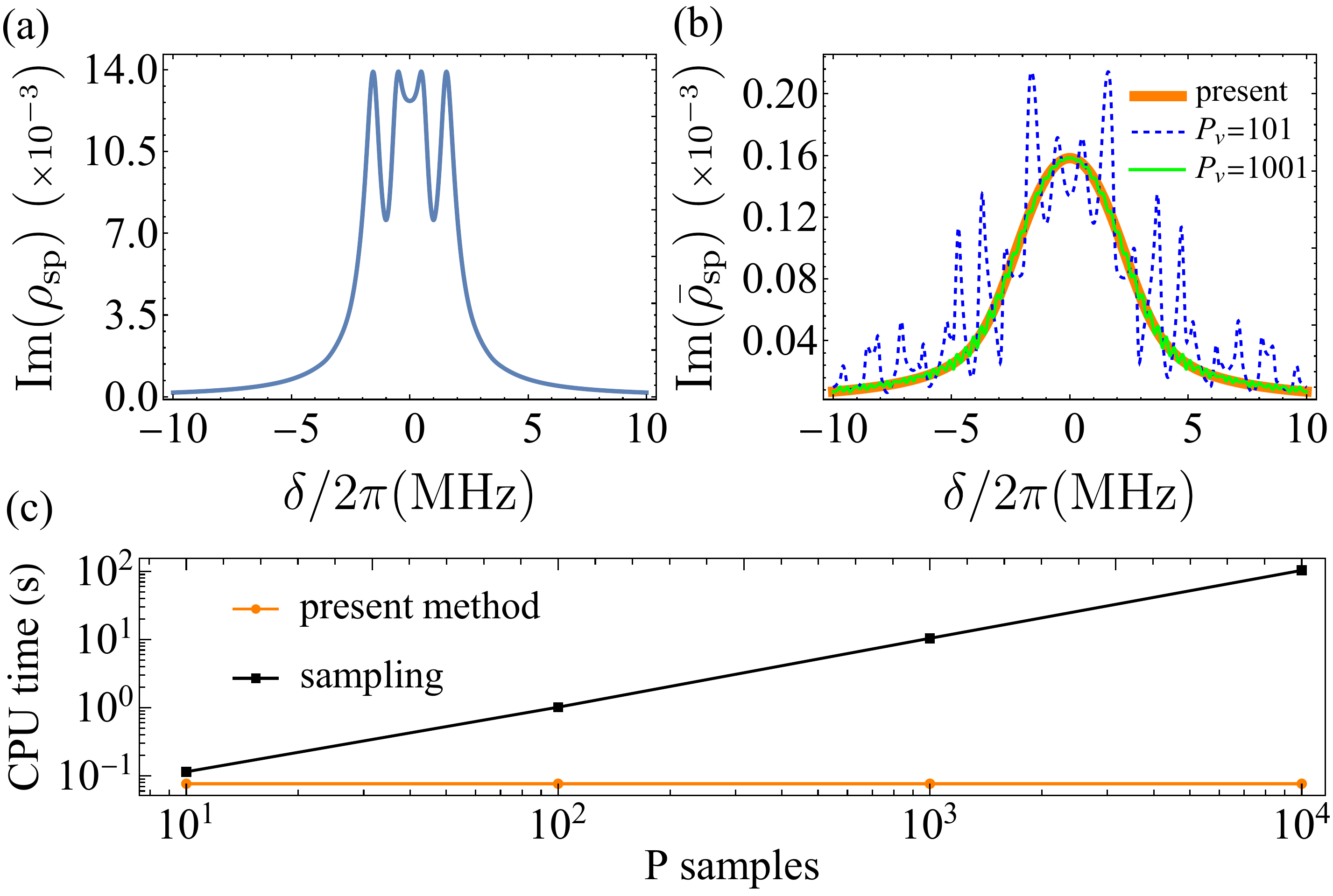}
   \caption{ (a) The peak-to-peak amplitude of ${\rm Im}\left[\rho_{\rm sp}(t)\right]$ versus the two-photon detuning $\delta$ at $v=0$. (b) The Doppler averaged signal of (a) using the present method (solid orange) and inverse transform sampling \cite{InverseSample} of $P_v=101$ (dashed blue) and $P_v=1001$ (solid black) velocities. (c) CPU time to compute the Doppler averaged signal for a given $\delta$ versus the number of velocity samples $P$. The system parameters are: $\Omega_1=2\pi \times 2 \ {\rm MHz},$ $\Omega_3=2\pi \times 1 \ {\rm MHz}$ and $\Gamma=2\pi \times 6 \ {\rm MHz}$. The modulation amplitude and frequency are $\Omega_2=2\pi \times 1 \ {\rm MHz}$ and $f=1 \ {\rm MHz}$. $k_1/2\pi=1/(0.78 \ \rm{\mu m})$ and $k_2/2\pi=-1/(1.248 \ \rm{\mu m})$. The detuning is set in the EIT resonance regime $\Delta=0$. }
    \label{fig:rho_sp_t}
\end{figure}

\subsection{Time-periodic perturbation}

The steady-state behavior of a system with harmonic perturbation, $\Omega_2(t)=\Omega_2 \cos(\omega t)$ in our case, can be handled with Floquet analysis \cite{QuantumMasterPro}. In the Floquet approach, the ansatz for $\bm{\rho}(t)$ is a Fourier decomposition in the harmonics of the modulating frequency $\omega$: $\bm{\rho}(t)=\sum_{m=-N_f}^{N_f} \bm{\rho}_m \exp(i m \omega t)$. A solution is obtained once all $\bm{\rho}_m$ are found that satisfy the equations of motion. The sum above yields the exact solution when all the harmonics are included, i.e., $N_f=\infty$. Practically, the number of higher-order harmonics is restricted to a finite integer $N_f$, which depends on the specific problem. The Liouville superoperator, constructed from $H$ and $L$, can be decomposed into time-independent and -dependent parts as $\mathcal{L}=\mathcal{L}'+\cos(\omega t) \mathcal{L}''$. Using $\cos(\omega t)=(e^{i\omega t}+e^{-i\omega t})/2$, and plugging the Floquet ansatz into the master equation [Eq. \eqref{QME}], we get the following equations relating the various harmonic components $\bm{\rho}_m$:
\begin{equation}\label{floquet_relations}
(\mathcal{L}' - im \omega) \bm{\rho}_m +\dfrac{\mathcal{L}''}{2}\bm{\rho}_{m+1}+ \dfrac{\mathcal{L}''}{2}\bm{\rho}_{m-1}=0
\end{equation}
If we construct a larger Floquet vector made from the harmonic components, $\bm{\rho}_F=(\bm{\rho}_{N_f},\bm{\rho}_{N_f-1},...,\bm{\rho}_{-N_f})^{\rm T}$, then this can be rewritten as:
\begin{equation}
\mathcal{L}_F \bm{\rho}_F=0,
\end{equation}
where the Floquet matrix $\mathcal{L}_F$ is explicitly given by [Eq. \eqref{floquet_relations}]
\begin{align}
&\mathcal{L}_F = \nonumber \\ & 
\begin{bmatrix}
  . & . & . & . & . & . & . & . & .  \\
  . & \mathcal{L}' - 2i \omega & \mathcal{L}''/2 & 0 & 0 & 0 & 0 & . & .  \\
  . & \mathcal{L}''/2 & \mathcal{L}' - i \omega & \mathcal{L}''/2 & 0 & 0 & 0 & . & .  \\
  . & 0 & \mathcal{L}''/2 & \mathcal{L}' & \mathcal{L}''/2 & 0 & 0 & . & .  \\
  . & 0 & 0 & \mathcal{L}''/2 & \mathcal{L}' + i \omega & \mathcal{L}''/2 & 0 & . & . \\
  . & 0 & 0 & 0 & \mathcal{L}''/2 & \mathcal{L}' + 2i \omega & \mathcal{L}''/2 & . & .\\
  . & . & . & . & . & . & . & . & .  \\
  . & . & . & . & . & . & . & . & .  \\
\end{bmatrix}
\end{align}
Therefore, the Floquet ansatz has transformed the problem into a time-independent one, where the nullspace of $\mathcal{L}_F$ gives $\bm{\rho}_F$, which in turn gives $\rho(t)$. Here, we shall include up to the $N_f=4$ harmonic.

As before, for $v \neq 0$, the detunings are Doppler shifted as $\Delta \rightarrow \Delta-k_1 v$ and $\delta \rightarrow \delta-k_2 v$. This changes the Hamiltonian and the corresponding Floquet matrix for every $v$. \rsub{The signal ${\rm Im}\left[\rho_{\rm sp}(t)\right]$ oscillates harmonically at steady state  due to the time modulation of the control field. In Fig. \ref{fig:rho_sp_t}(a) we show the peak-to-peak amplitude of ${\rm Im}\left[\rho_{\rm sp}(t)\right]$ versus the detuning $\delta$ at $v=0$. The actual observed signal is the coherence averaged over the Maxwell-Boltzmann distribution $P(v)$ with a thermal velocity $v_{\rm th}$. Computing the Doppler broadening by brute force would involve sampling the steady state over $v$ for hundreds or thousands of velocity classes and approximating the average by a Riemann sum.}  

To compute Doppler broadening by the present approach, we proceed as follows. We find $\bm{\rho}_0$ at $v=0$ and construct $\mathcal{L}_F^{-}$. Next, we construct $\mathcal{L}_1$ by taking the Doppler shifted component of $\mathcal{L}_F$, i.e., $\mathcal{L}_1=d\mathcal{L}_F(\Delta-k_1 v,\delta-k_2 v)/dv$. Then we diagonalize $\mathcal{L}_F^{-} \mathcal{L}_1$. Finally, the exact Doppler averaged signal will be given by Eq. \eqref{rho_gauss_1D}. \rsub{ Fig. \ref{fig:rho_sp_t}(b) shows the Doppler averaged signal versus $\delta$ using the present method (orange), without any sampling. For comparison, we Doppler averaged the signal using $P=101$ (blue dashed) and $P=1001$ (green) velocity classes, where we used the inverse transform sampling approach, given its efficiency \cite{InverseSample}. One sees that large $P=10^3$ is needed for the approximate average to start converging to the exact values, which still suffers from numerical artifacts such as oscillations due to finite sampling. The Doppler averaged signal due to the present method, on the other hand, is smooth and does not suffer from any numerical artifacts, given that it is the exact average. Fig. \ref{fig:rho_sp_t}(c) shows the time taken to do the averaging using the present (orange) and the sampling (black) approaches versus $P$. While the time for the sampling approach scales as $O(P)$, the present method is independent of $P$, since it only requires diagonalizing $\mathcal{L}_F^{-} \mathcal{L}_1$. Given that at least $P=10^3-10^4$ would be needed for the sampling approach to start converging to the same accuracy as the present method, the present method achieves between two and three orders of magnitude relative speedup, while also being exact. For a Python-based implementation of the present analytical Doppler method, see the open-source package RydIQule Version 2 \cite{miller2025rydiqule,rydiqule_doppler_demo2025}.} 

\section{Discussion}\label{sec:discuss}

We have presented a general non-perturbative approach to compute the exact \rsub{steady} state of open quantum systems under time-independent or time-periodic perturbations. This work goes beyond previous perturbative frameworks \cite{perturb_theory_2014, Perturbation_infinite_resum_2016}, where the present exact solution exists even when a corresponding perturbative solution does not. The main result is a derivation of the propagator $G_v$ that generates the entire dependence of the steady state on the perturbation, just using \rsub{a single diagonalization}. The main technical result that enabled the construction of $G_v$ is the use of a particular generalized inverse, the Drazin inverse $\mathcal{L}_0^-$ [Eq. \eqref{L0_minus}]. This generalized inverse naturally retains the normalization of the original unperturbed steady state via Eq. \eqref{generalized_property}, where $\bm{L}_0={\rm vec}(\mathbb{1})$ and \rsub{the unperturbed steady state} $\bm{R}_0=\bm{\rho}_0$ appear explicitly. This allows for efficient generation of the steady state, as well as exact analytic operations, without sampling or discretization. A particularly useful feature of this approach is that it works even if the steady state is only available numerically. We remark that it is in fact possible to construct a propagator $G_v$ using a different generalized inverse, e.g., the Moore-Penrose inverse, which would also generate the exact state state via $\bm{\rho}_v=G_v \bm{\rho}_0$. However, in this case, the steady state would need to be renormalized for every distinct $v$, and therefore $\bm{\rho}_v$ would be a non-trivial function of $v$. This would make it challenging to do any analytic operations on $\bm{\rho}_v$ efficiently or exactly. Therefore, we conclude that the present generalized inverse [Eq. \eqref{L0_minus}] is of central importance in the present theory.

The derivation of the propagator $G_v$ relied on general assumptions, namely probability conservation during the time evolution, a unique steady state (see Ref. \cite{steady_state_cond_1976} for sufficient conditions on uniqueness), and that $\mathcal{L}$ and $\mathcal{L}_0^-\mathcal{L}_1$ admit eigendecompositions. For some systems, $\mathcal{L}$ does not admit an eigendecomposition \cite{L_eigen2023, QuantumMasterPro}. We leave it for future work to rigorously classify which class of open quantum systems satisfy the above assumptions, and hence can be handled with the present method. An interesting future direction is to extend the present approach to systems that do not admit such an eigendecomposition. 

We have applied the present method on three non-trivial quantum systems, showing that it agrees with both exact and numerical solutions. In addition to being exact, we have demonstrated the method's utility for efficient numerical simulation, achieving a speedup of one to several orders of magnitude. The present method is expected to offer speedup for problems that require large sampling, when the computational time for diagonalizing the unperturbed system is much less than the time taken to repeatedly solve for the nullspace for different values of the perturbation. In particular, the present approach is well-positioned to offer speedup for problems involving ensemble averaging, e.g., inhomogeneous broadening calculations, because it completely avoids sampling. 

\section*{Acknowledgments}
We thank Anirudh Yadav for valuable discussions. This work was partially supported by the National Science Foundation (Grants No. GOALI PHY-1912543 and No. 2016136 for the QLCI Hybrid Quantum Architectures and Networks).

\section*{Data availability}
The data that support the findings of this article are openly available \cite{nagib_exact_2025}.


\bibliography{doppler}

\begin{thebibliography}{57}%
\makeatletter
\providecommand \@ifxundefined [1]{%
 \@ifx{#1\undefined}
}%
\providecommand \@ifnum [1]{%
 \ifnum #1\expandafter \@firstoftwo
 \else \expandafter \@secondoftwo
 \fi
}%
\providecommand \@ifx [1]{%
 \ifx #1\expandafter \@firstoftwo
 \else \expandafter \@secondoftwo
 \fi
}%
\providecommand \natexlab [1]{#1}%
\providecommand \enquote  [1]{``#1''}%
\providecommand \bibnamefont  [1]{#1}%
\providecommand \bibfnamefont [1]{#1}%
\providecommand \citenamefont [1]{#1}%
\providecommand \href@noop [0]{\@secondoftwo}%
\providecommand \href [0]{\begingroup \@sanitize@url \@href}%
\providecommand \@href[1]{\@@startlink{#1}\@@href}%
\providecommand \@@href[1]{\endgroup#1\@@endlink}%
\providecommand \@sanitize@url [0]{\catcode `\\12\catcode `\$12\catcode `\&12\catcode `\#12\catcode `\^12\catcode `\_12\catcode `\%12\relax}%
\providecommand \@@startlink[1]{}%
\providecommand \@@endlink[0]{}%
\providecommand \url  [0]{\begingroup\@sanitize@url \@url }%
\providecommand \@url [1]{\endgroup\@href {#1}{\urlprefix }}%
\providecommand \urlprefix  [0]{URL }%
\providecommand \Eprint [0]{\href }%
\providecommand \doibase [0]{https://doi.org/}%
\providecommand \selectlanguage [0]{\@gobble}%
\providecommand \bibinfo  [0]{\@secondoftwo}%
\providecommand \bibfield  [0]{\@secondoftwo}%
\providecommand \translation [1]{[#1]}%
\providecommand \BibitemOpen [0]{}%
\providecommand \bibitemStop [0]{}%
\providecommand \bibitemNoStop [0]{.\EOS\space}%
\providecommand \EOS [0]{\spacefactor3000\relax}%
\providecommand \BibitemShut  [1]{\csname bibitem#1\endcsname}%
\let\auto@bib@innerbib\@empty
\bibitem [{\citenamefont {Campaioli}\ \emph {et~al.}(2024)\citenamefont {Campaioli}, \citenamefont {Cole},\ and\ \citenamefont {Hapuarachchi}}]{QuantumMasterPro}%
  \BibitemOpen
  \bibfield  {author} {\bibinfo {author} {\bibfnamefont {F.}~\bibnamefont {Campaioli}}, \bibinfo {author} {\bibfnamefont {J.~H.}\ \bibnamefont {Cole}},\ and\ \bibinfo {author} {\bibfnamefont {H.}~\bibnamefont {Hapuarachchi}},\ }\bibfield  {title} {\bibinfo {title} {{Quantum Master Equations: Tips and Tricks for Quantum Optics, Quantum Computing, and Beyond}},\ }\href {https://doi.org/10.1103/PRXQuantum.5.020202} {\bibfield  {journal} {\bibinfo  {journal} {PRX Quantum}\ }\textbf {\bibinfo {volume} {5}},\ \bibinfo {pages} {020202} (\bibinfo {year} {2024})}\BibitemShut {NoStop}%
\bibitem [{\citenamefont {Nation}(2015)}]{Review_2015}%
  \BibitemOpen
  \bibfield  {author} {\bibinfo {author} {\bibfnamefont {P.~D.}\ \bibnamefont {Nation}},\ }\href {https://arxiv.org/abs/1504.06768} {\bibinfo {title} {Steady-state solution methods for open quantum optical systems}} (\bibinfo {year} {2015}),\ \Eprint {https://arxiv.org/abs/1504.06768} {arXiv:1504.06768 [quant-ph]} \BibitemShut {NoStop}%
\bibitem [{\citenamefont {Cui}\ \emph {et~al.}(2015)\citenamefont {Cui}, \citenamefont {Cirac},\ and\ \citenamefont {Ba\~nuls}}]{Variational_2015}%
  \BibitemOpen
  \bibfield  {author} {\bibinfo {author} {\bibfnamefont {J.}~\bibnamefont {Cui}}, \bibinfo {author} {\bibfnamefont {J.~I.}\ \bibnamefont {Cirac}},\ and\ \bibinfo {author} {\bibfnamefont {M.~C.}\ \bibnamefont {Ba\~nuls}},\ }\bibfield  {title} {\bibinfo {title} {{Variational Matrix Product Operators for the Steady State of Dissipative Quantum Systems}},\ }\href {https://doi.org/10.1103/PhysRevLett.114.220601} {\bibfield  {journal} {\bibinfo  {journal} {Phys. Rev. Lett.}\ }\textbf {\bibinfo {volume} {114}},\ \bibinfo {pages} {220601} (\bibinfo {year} {2015})}\BibitemShut {NoStop}%
\bibitem [{\citenamefont {Melo}\ \emph {et~al.}(2025)\citenamefont {Melo}, \citenamefont {Beugnot},\ and\ \citenamefont {Minganti}}]{VPT2025}%
  \BibitemOpen
  \bibfield  {author} {\bibinfo {author} {\bibfnamefont {A.}~\bibnamefont {Melo}}, \bibinfo {author} {\bibfnamefont {G.}~\bibnamefont {Beugnot}},\ and\ \bibinfo {author} {\bibfnamefont {F.}~\bibnamefont {Minganti}},\ }\href {https://arxiv.org/abs/2504.00085} {\bibinfo {title} {{Variational Perturbation Theory in Open Quantum Systems for Efficient Steady State Computation}}} (\bibinfo {year} {2025}),\ \Eprint {https://arxiv.org/abs/2504.00085} {arXiv:2504.00085 [quant-ph]} \BibitemShut {NoStop}%
\bibitem [{\citenamefont {Vicentini}\ \emph {et~al.}(2019{\natexlab{a}})\citenamefont {Vicentini}, \citenamefont {Biella}, \citenamefont {Regnault},\ and\ \citenamefont {Ciuti}}]{Neural_2019}%
  \BibitemOpen
  \bibfield  {author} {\bibinfo {author} {\bibfnamefont {F.}~\bibnamefont {Vicentini}}, \bibinfo {author} {\bibfnamefont {A.}~\bibnamefont {Biella}}, \bibinfo {author} {\bibfnamefont {N.}~\bibnamefont {Regnault}},\ and\ \bibinfo {author} {\bibfnamefont {C.}~\bibnamefont {Ciuti}},\ }\bibfield  {title} {\bibinfo {title} {{Variational Neural-Network Ansatz for Steady States in Open Quantum Systems}},\ }\href {https://doi.org/10.1103/PhysRevLett.122.250503} {\bibfield  {journal} {\bibinfo  {journal} {Phys. Rev. Lett.}\ }\textbf {\bibinfo {volume} {122}},\ \bibinfo {pages} {250503} (\bibinfo {year} {2019}{\natexlab{a}})}\BibitemShut {NoStop}%
\bibitem [{\citenamefont {Dalibard}\ \emph {et~al.}(1992)\citenamefont {Dalibard}, \citenamefont {Castin},\ and\ \citenamefont {M\o{}lmer}}]{Quantum_trajectory_1992}%
  \BibitemOpen
  \bibfield  {author} {\bibinfo {author} {\bibfnamefont {J.}~\bibnamefont {Dalibard}}, \bibinfo {author} {\bibfnamefont {Y.}~\bibnamefont {Castin}},\ and\ \bibinfo {author} {\bibfnamefont {K.}~\bibnamefont {M\o{}lmer}},\ }\bibfield  {title} {\bibinfo {title} {Wave-function approach to dissipative processes in quantum optics},\ }\href {https://doi.org/10.1103/PhysRevLett.68.580} {\bibfield  {journal} {\bibinfo  {journal} {Phys. Rev. Lett.}\ }\textbf {\bibinfo {volume} {68}},\ \bibinfo {pages} {580} (\bibinfo {year} {1992})}\BibitemShut {NoStop}%
\bibitem [{\citenamefont {Vicentini}\ \emph {et~al.}(2019{\natexlab{b}})\citenamefont {Vicentini}, \citenamefont {Minganti}, \citenamefont {Biella}, \citenamefont {Orso},\ and\ \citenamefont {Ciuti}}]{Trajectory_static_disorder_2010}%
  \BibitemOpen
  \bibfield  {author} {\bibinfo {author} {\bibfnamefont {F.}~\bibnamefont {Vicentini}}, \bibinfo {author} {\bibfnamefont {F.}~\bibnamefont {Minganti}}, \bibinfo {author} {\bibfnamefont {A.}~\bibnamefont {Biella}}, \bibinfo {author} {\bibfnamefont {G.}~\bibnamefont {Orso}},\ and\ \bibinfo {author} {\bibfnamefont {C.}~\bibnamefont {Ciuti}},\ }\bibfield  {title} {\bibinfo {title} {Optimal stochastic unraveling of disordered open quantum systems: {Application} to driven-dissipative photonic lattices},\ }\href {https://doi.org/10.1103/PhysRevA.99.032115} {\bibfield  {journal} {\bibinfo  {journal} {Phys. Rev. A}\ }\textbf {\bibinfo {volume} {99}},\ \bibinfo {pages} {032115} (\bibinfo {year} {2019}{\natexlab{b}})}\BibitemShut {NoStop}%
\bibitem [{\citenamefont {Verstraete}\ \emph {et~al.}(2004)\citenamefont {Verstraete}, \citenamefont {Garc\'{\i}a-Ripoll},\ and\ \citenamefont {Cirac}}]{MPDO}%
  \BibitemOpen
  \bibfield  {author} {\bibinfo {author} {\bibfnamefont {F.}~\bibnamefont {Verstraete}}, \bibinfo {author} {\bibfnamefont {J.~J.}\ \bibnamefont {Garc\'{\i}a-Ripoll}},\ and\ \bibinfo {author} {\bibfnamefont {J.~I.}\ \bibnamefont {Cirac}},\ }\bibfield  {title} {\bibinfo {title} {{Matrix Product Density Operators: Simulation of Finite-Temperature and Dissipative Systems}},\ }\href {https://doi.org/10.1103/PhysRevLett.93.207204} {\bibfield  {journal} {\bibinfo  {journal} {Phys. Rev. Lett.}\ }\textbf {\bibinfo {volume} {93}},\ \bibinfo {pages} {207204} (\bibinfo {year} {2004})}\BibitemShut {NoStop}%
\bibitem [{\citenamefont {Nation}\ \emph {et~al.}(2015)\citenamefont {Nation}, \citenamefont {Johansson}, \citenamefont {Blencowe},\ and\ \citenamefont {Rimberg}}]{Iter_2015}%
  \BibitemOpen
  \bibfield  {author} {\bibinfo {author} {\bibfnamefont {P.~D.}\ \bibnamefont {Nation}}, \bibinfo {author} {\bibfnamefont {J.~R.}\ \bibnamefont {Johansson}}, \bibinfo {author} {\bibfnamefont {M.~P.}\ \bibnamefont {Blencowe}},\ and\ \bibinfo {author} {\bibfnamefont {A.~J.}\ \bibnamefont {Rimberg}},\ }\bibfield  {title} {\bibinfo {title} {Iterative solutions to the steady-state density matrix for optomechanical systems},\ }\href {https://doi.org/10.1103/PhysRevE.91.013307} {\bibfield  {journal} {\bibinfo  {journal} {Phys. Rev. E}\ }\textbf {\bibinfo {volume} {91}},\ \bibinfo {pages} {013307} (\bibinfo {year} {2015})}\BibitemShut {NoStop}%
\bibitem [{\citenamefont {Kaspar}\ and\ \citenamefont {Thoss}(2021)}]{HEOM_2021}%
  \BibitemOpen
  \bibfield  {author} {\bibinfo {author} {\bibfnamefont {C.}~\bibnamefont {Kaspar}}\ and\ \bibinfo {author} {\bibfnamefont {M.}~\bibnamefont {Thoss}},\ }\bibfield  {title} {\bibinfo {title} {{Efficient Steady-State Solver for the Hierarchical Equations of Motion Approach: Formulation and Application to Charge Transport through Nanosystems}},\ }\href {https://doi.org/10.1021/acs.jpca.1c02863} {\bibfield  {journal} {\bibinfo  {journal} {The Journal of Physical Chemistry A}\ }\textbf {\bibinfo {volume} {125}},\ \bibinfo {pages} {5190–5200} (\bibinfo {year} {2021})}\BibitemShut {NoStop}%
\bibitem [{\citenamefont {Deffner}\ and\ \citenamefont {Lutz}(2011)}]{Entropy_2011}%
  \BibitemOpen
  \bibfield  {author} {\bibinfo {author} {\bibfnamefont {S.}~\bibnamefont {Deffner}}\ and\ \bibinfo {author} {\bibfnamefont {E.}~\bibnamefont {Lutz}},\ }\bibfield  {title} {\bibinfo {title} {{Nonequilibrium Entropy Production for Open Quantum Systems}},\ }\href {https://doi.org/10.1103/PhysRevLett.107.140404} {\bibfield  {journal} {\bibinfo  {journal} {Phys. Rev. Lett.}\ }\textbf {\bibinfo {volume} {107}},\ \bibinfo {pages} {140404} (\bibinfo {year} {2011})}\BibitemShut {NoStop}%
\bibitem [{\citenamefont {Ajisaka}\ \emph {et~al.}(2012)\citenamefont {Ajisaka}, \citenamefont {Barra}, \citenamefont {Mej\'{\i}a-Monasterio},\ and\ \citenamefont {Prosen}}]{NESS_mesoscopic_2012}%
  \BibitemOpen
  \bibfield  {author} {\bibinfo {author} {\bibfnamefont {S.}~\bibnamefont {Ajisaka}}, \bibinfo {author} {\bibfnamefont {F.}~\bibnamefont {Barra}}, \bibinfo {author} {\bibfnamefont {C.}~\bibnamefont {Mej\'{\i}a-Monasterio}},\ and\ \bibinfo {author} {\bibfnamefont {T.~c.~v.}\ \bibnamefont {Prosen}},\ }\bibfield  {title} {\bibinfo {title} {{Nonequlibrium particle and energy currents in quantum chains connected to mesoscopic Fermi reservoirs}},\ }\href {https://doi.org/10.1103/PhysRevB.86.125111} {\bibfield  {journal} {\bibinfo  {journal} {Phys. Rev. B}\ }\textbf {\bibinfo {volume} {86}},\ \bibinfo {pages} {125111} (\bibinfo {year} {2012})}\BibitemShut {NoStop}%
\bibitem [{\citenamefont {Esposito}\ \emph {et~al.}(2015)\citenamefont {Esposito}, \citenamefont {Ochoa},\ and\ \citenamefont {Galperin}}]{NESS_quantum_thermo_2015}%
  \BibitemOpen
  \bibfield  {author} {\bibinfo {author} {\bibfnamefont {M.}~\bibnamefont {Esposito}}, \bibinfo {author} {\bibfnamefont {M.~A.}\ \bibnamefont {Ochoa}},\ and\ \bibinfo {author} {\bibfnamefont {M.}~\bibnamefont {Galperin}},\ }\bibfield  {title} {\bibinfo {title} {{Quantum Thermodynamics: A Nonequilibrium Green's Function Approach}},\ }\href {https://doi.org/10.1103/PhysRevLett.114.080602} {\bibfield  {journal} {\bibinfo  {journal} {Phys. Rev. Lett.}\ }\textbf {\bibinfo {volume} {114}},\ \bibinfo {pages} {080602} (\bibinfo {year} {2015})}\BibitemShut {NoStop}%
\bibitem [{\citenamefont {Topp}\ \emph {et~al.}(2015)\citenamefont {Topp}, \citenamefont {Brandes},\ and\ \citenamefont {Schaller}}]{Topp_2015}%
  \BibitemOpen
  \bibfield  {author} {\bibinfo {author} {\bibfnamefont {G.~E.}\ \bibnamefont {Topp}}, \bibinfo {author} {\bibfnamefont {T.}~\bibnamefont {Brandes}},\ and\ \bibinfo {author} {\bibfnamefont {G.}~\bibnamefont {Schaller}},\ }\bibfield  {title} {\bibinfo {title} {{Steady-state thermodynamics of non-interacting transport beyond weak coupling}},\ }\href {https://doi.org/10.1209/0295-5075/110/67003} {\bibfield  {journal} {\bibinfo  {journal} {EPL (Europhysics Letters)}\ }\textbf {\bibinfo {volume} {110}},\ \bibinfo {pages} {67003} (\bibinfo {year} {2015})}\BibitemShut {NoStop}%
\bibitem [{\citenamefont {Ness}(2017)}]{Ness_2017}%
  \BibitemOpen
  \bibfield  {author} {\bibinfo {author} {\bibfnamefont {H.}~\bibnamefont {Ness}},\ }\bibfield  {title} {\bibinfo {title} {{Nonequilibrium Thermodynamics and Steady State Density Matrix for Quantum Open Systems}},\ }\href {https://doi.org/10.3390/e19040158} {\bibfield  {journal} {\bibinfo  {journal} {Entropy}\ }\textbf {\bibinfo {volume} {19}},\ \bibinfo {pages} {158} (\bibinfo {year} {2017})}\BibitemShut {NoStop}%
\bibitem [{\citenamefont {Weimer}\ \emph {et~al.}(2021)\citenamefont {Weimer}, \citenamefont {Kshetrimayum},\ and\ \citenamefont {Or\'us}}]{Many_body_review_2021}%
  \BibitemOpen
  \bibfield  {author} {\bibinfo {author} {\bibfnamefont {H.}~\bibnamefont {Weimer}}, \bibinfo {author} {\bibfnamefont {A.}~\bibnamefont {Kshetrimayum}},\ and\ \bibinfo {author} {\bibfnamefont {R.}~\bibnamefont {Or\'us}},\ }\bibfield  {title} {\bibinfo {title} {Simulation methods for open quantum many-body systems},\ }\href {https://doi.org/10.1103/RevModPhys.93.015008} {\bibfield  {journal} {\bibinfo  {journal} {Rev. Mod. Phys.}\ }\textbf {\bibinfo {volume} {93}},\ \bibinfo {pages} {015008} (\bibinfo {year} {2021})}\BibitemShut {NoStop}%
\bibitem [{\citenamefont {Johansson}\ \emph {et~al.}(2012)\citenamefont {Johansson}, \citenamefont {Nation},\ and\ \citenamefont {Nori}}]{QuTiP}%
  \BibitemOpen
  \bibfield  {author} {\bibinfo {author} {\bibfnamefont {J.}~\bibnamefont {Johansson}}, \bibinfo {author} {\bibfnamefont {P.}~\bibnamefont {Nation}},\ and\ \bibinfo {author} {\bibfnamefont {F.}~\bibnamefont {Nori}},\ }\bibfield  {title} {\bibinfo {title} {Qutip: {An} open-source {Python} framework for the dynamics of open quantum systems},\ }\href {https://doi.org/10.1016/j.cpc.2012.02.021} {\bibfield  {journal} {\bibinfo  {journal} {Computer Physics Communications}\ }\textbf {\bibinfo {volume} {183}},\ \bibinfo {pages} {1760–1772} (\bibinfo {year} {2012})}\BibitemShut {NoStop}%
\bibitem [{\citenamefont {Chen}\ and\ \citenamefont {Lidar}(2022)}]{HOQST}%
  \BibitemOpen
  \bibfield  {author} {\bibinfo {author} {\bibfnamefont {H.}~\bibnamefont {Chen}}\ and\ \bibinfo {author} {\bibfnamefont {D.~A.}\ \bibnamefont {Lidar}},\ }\bibfield  {title} {\bibinfo {title} {Hamiltonian open quantum system toolkit},\ }\bibfield  {journal} {\bibinfo  {journal} {Communications Physics}\ }\textbf {\bibinfo {volume} {5}},\ \href {https://doi.org/10.1038/s42005-022-00887-2} {10.1038/s42005-022-00887-2} (\bibinfo {year} {2022})\BibitemShut {NoStop}%
\bibitem [{\citenamefont {Hogben}\ \emph {et~al.}(2011)\citenamefont {Hogben}, \citenamefont {Krzystyniak}, \citenamefont {Charnock}, \citenamefont {Hore},\ and\ \citenamefont {Kuprov}}]{SPINACH}%
  \BibitemOpen
  \bibfield  {author} {\bibinfo {author} {\bibfnamefont {H.}~\bibnamefont {Hogben}}, \bibinfo {author} {\bibfnamefont {M.}~\bibnamefont {Krzystyniak}}, \bibinfo {author} {\bibfnamefont {G.}~\bibnamefont {Charnock}}, \bibinfo {author} {\bibfnamefont {P.}~\bibnamefont {Hore}},\ and\ \bibinfo {author} {\bibfnamefont {I.}~\bibnamefont {Kuprov}},\ }\bibfield  {title} {\bibinfo {title} {Spinach – {A} software library for simulation of spin dynamics in large spin systems},\ }\href {https://doi.org/https://doi.org/10.1016/j.jmr.2010.11.008} {\bibfield  {journal} {\bibinfo  {journal} {Journal of Magnetic Resonance}\ }\textbf {\bibinfo {volume} {208}},\ \bibinfo {pages} {179} (\bibinfo {year} {2011})}\BibitemShut {NoStop}%
\bibitem [{\citenamefont {Krämer}\ \emph {et~al.}(2018)\citenamefont {Krämer}, \citenamefont {Plankensteiner}, \citenamefont {Ostermann},\ and\ \citenamefont {Ritsch}}]{Julia}%
  \BibitemOpen
  \bibfield  {author} {\bibinfo {author} {\bibfnamefont {S.}~\bibnamefont {Krämer}}, \bibinfo {author} {\bibfnamefont {D.}~\bibnamefont {Plankensteiner}}, \bibinfo {author} {\bibfnamefont {L.}~\bibnamefont {Ostermann}},\ and\ \bibinfo {author} {\bibfnamefont {H.}~\bibnamefont {Ritsch}},\ }\bibfield  {title} {\bibinfo {title} {Quantumoptics.jl: {A Julia} framework for simulating open quantum systems},\ }\href {https://doi.org/10.1016/j.cpc.2018.02.004} {\bibfield  {journal} {\bibinfo  {journal} {Computer Physics Communications}\ }\textbf {\bibinfo {volume} {227}},\ \bibinfo {pages} {109–116} (\bibinfo {year} {2018})}\BibitemShut {NoStop}%
\bibitem [{\citenamefont {Miller}\ \emph {et~al.}(2024)\citenamefont {Miller}, \citenamefont {Meyer}, \citenamefont {Virtanen}, \citenamefont {O'Brien},\ and\ \citenamefont {Cox}}]{RydIQule}%
  \BibitemOpen
  \bibfield  {author} {\bibinfo {author} {\bibfnamefont {B.~N.}\ \bibnamefont {Miller}}, \bibinfo {author} {\bibfnamefont {D.~H.}\ \bibnamefont {Meyer}}, \bibinfo {author} {\bibfnamefont {T.}~\bibnamefont {Virtanen}}, \bibinfo {author} {\bibfnamefont {C.~M.}\ \bibnamefont {O'Brien}},\ and\ \bibinfo {author} {\bibfnamefont {K.~C.}\ \bibnamefont {Cox}},\ }\bibfield  {title} {\bibinfo {title} {Rydiqule: {A} graph-based paradigm for modeling {Rydberg} and atomic sensors},\ }\href {https://doi.org/https://doi.org/10.1016/j.cpc.2023.108952} {\bibfield  {journal} {\bibinfo  {journal} {Computer Physics Communications}\ }\textbf {\bibinfo {volume} {294}},\ \bibinfo {pages} {108952} (\bibinfo {year} {2024})}\BibitemShut {NoStop}%
\bibitem [{\citenamefont {Finkelstein}\ \emph {et~al.}(2023)\citenamefont {Finkelstein}, \citenamefont {Bali}, \citenamefont {Firstenberg},\ and\ \citenamefont {Novikova}}]{practicalEIT}%
  \BibitemOpen
  \bibfield  {author} {\bibinfo {author} {\bibfnamefont {R.}~\bibnamefont {Finkelstein}}, \bibinfo {author} {\bibfnamefont {S.}~\bibnamefont {Bali}}, \bibinfo {author} {\bibfnamefont {O.}~\bibnamefont {Firstenberg}},\ and\ \bibinfo {author} {\bibfnamefont {I.}~\bibnamefont {Novikova}},\ }\bibfield  {title} {\bibinfo {title} {A practical guide to electromagnetically induced transparency in atomic vapor},\ }\href {https://doi.org/10.1088/1367-2630/acbc40} {\bibfield  {journal} {\bibinfo  {journal} {New Journal of Physics}\ }\textbf {\bibinfo {volume} {25}},\ \bibinfo {pages} {035001} (\bibinfo {year} {2023})}\BibitemShut {NoStop}%
\bibitem [{\citenamefont {Zanardi}\ and\ \citenamefont {Campos~Venuti}(2015)}]{dissipation_projected_dynamics_2015}%
  \BibitemOpen
  \bibfield  {author} {\bibinfo {author} {\bibfnamefont {P.}~\bibnamefont {Zanardi}}\ and\ \bibinfo {author} {\bibfnamefont {L.}~\bibnamefont {Campos~Venuti}},\ }\bibfield  {title} {\bibinfo {title} {Geometry, robustness, and emerging unitarity in dissipation-projected dynamics},\ }\href {https://doi.org/10.1103/PhysRevA.91.052324} {\bibfield  {journal} {\bibinfo  {journal} {Phys. Rev. A}\ }\textbf {\bibinfo {volume} {91}},\ \bibinfo {pages} {052324} (\bibinfo {year} {2015})}\BibitemShut {NoStop}%
\bibitem [{\citenamefont {Benatti}\ \emph {et~al.}(2011)\citenamefont {Benatti}, \citenamefont {Nagy},\ and\ \citenamefont {Narnhofer}}]{perturb_theory_steady_2011}%
  \BibitemOpen
  \bibfield  {author} {\bibinfo {author} {\bibfnamefont {F.}~\bibnamefont {Benatti}}, \bibinfo {author} {\bibfnamefont {A.}~\bibnamefont {Nagy}},\ and\ \bibinfo {author} {\bibfnamefont {H.}~\bibnamefont {Narnhofer}},\ }\bibfield  {title} {\bibinfo {title} {Asymptotic entanglement and {Lindblad} dynamics: a perturbative approach},\ }\href {https://doi.org/10.1088/1751-8113/44/15/155303} {\bibfield  {journal} {\bibinfo  {journal} {Journal of Physics A: Mathematical and Theoretical}\ }\textbf {\bibinfo {volume} {44}},\ \bibinfo {pages} {155303} (\bibinfo {year} {2011})}\BibitemShut {NoStop}%
\bibitem [{\citenamefont {del Valle}\ and\ \citenamefont {Hartmann}(2013)}]{perturb_theory_steady_2013}%
  \BibitemOpen
  \bibfield  {author} {\bibinfo {author} {\bibfnamefont {E.}~\bibnamefont {del Valle}}\ and\ \bibinfo {author} {\bibfnamefont {M.~J.}\ \bibnamefont {Hartmann}},\ }\bibfield  {title} {\bibinfo {title} {Correlator expansion approach to stationary states of weakly coupled cavity arrays},\ }\href {https://doi.org/10.1088/0953-4075/46/22/224023} {\bibfield  {journal} {\bibinfo  {journal} {Journal of Physics B: Atomic, Molecular and Optical Physics}\ }\textbf {\bibinfo {volume} {46}},\ \bibinfo {pages} {224023} (\bibinfo {year} {2013})}\BibitemShut {NoStop}%
\bibitem [{\citenamefont {Flindt}\ \emph {et~al.}(2008)\citenamefont {Flindt}, \citenamefont {Novotný}, \citenamefont {Braggio}, \citenamefont {Sassetti},\ and\ \citenamefont {Jauho}}]{Counting_stat_2008}%
  \BibitemOpen
  \bibfield  {author} {\bibinfo {author} {\bibfnamefont {C.}~\bibnamefont {Flindt}}, \bibinfo {author} {\bibfnamefont {T.}~\bibnamefont {Novotný}}, \bibinfo {author} {\bibfnamefont {A.}~\bibnamefont {Braggio}}, \bibinfo {author} {\bibfnamefont {M.}~\bibnamefont {Sassetti}},\ and\ \bibinfo {author} {\bibfnamefont {A.-P.}\ \bibnamefont {Jauho}},\ }\bibfield  {title} {\bibinfo {title} {{Counting Statistics of Non-Markovian Quantum Stochastic Processes}},\ }\bibfield  {journal} {\bibinfo  {journal} {Physical Review Letters}\ }\textbf {\bibinfo {volume} {100}},\ \href {https://doi.org/10.1103/physrevlett.100.150601} {10.1103/physrevlett.100.150601} (\bibinfo {year} {2008})\BibitemShut {NoStop}%
\bibitem [{\citenamefont {\ifmmode \check{Z}\else \v{Z}\fi{}nidari\ifmmode~\check{c}\else \v{c}\fi{}}(2015)}]{Perturb_fail_2015}%
  \BibitemOpen
  \bibfield  {author} {\bibinfo {author} {\bibfnamefont {M.}~\bibnamefont {\ifmmode \check{Z}\else \v{Z}\fi{}nidari\ifmmode~\check{c}\else \v{c}\fi{}}},\ }\bibfield  {title} {\bibinfo {title} {Relaxation times of dissipative many-body quantum systems},\ }\href {https://doi.org/10.1103/PhysRevE.92.042143} {\bibfield  {journal} {\bibinfo  {journal} {Phys. Rev. E}\ }\textbf {\bibinfo {volume} {92}},\ \bibinfo {pages} {042143} (\bibinfo {year} {2015})}\BibitemShut {NoStop}%
\bibitem [{\citenamefont {Li}\ \emph {et~al.}(2023)\citenamefont {Li}, \citenamefont {Wu}, \citenamefont {Zheng},\ and\ \citenamefont {Yi}}]{Many_body_perturb_2023}%
  \BibitemOpen
  \bibfield  {author} {\bibinfo {author} {\bibfnamefont {H.}~\bibnamefont {Li}}, \bibinfo {author} {\bibfnamefont {H.}~\bibnamefont {Wu}}, \bibinfo {author} {\bibfnamefont {W.}~\bibnamefont {Zheng}},\ and\ \bibinfo {author} {\bibfnamefont {W.}~\bibnamefont {Yi}},\ }\bibfield  {title} {\bibinfo {title} {{Many}-body non-{Hermitian} skin effect under dynamic gauge coupling},\ }\href {https://doi.org/10.1103/PhysRevResearch.5.033173} {\bibfield  {journal} {\bibinfo  {journal} {Phys. Rev. Res.}\ }\textbf {\bibinfo {volume} {5}},\ \bibinfo {pages} {033173} (\bibinfo {year} {2023})}\BibitemShut {NoStop}%
\bibitem [{\citenamefont {Shishkov}\ \emph {et~al.}(2020)\citenamefont {Shishkov}, \citenamefont {Andrianov}, \citenamefont {Pukhov}, \citenamefont {Vinogradov},\ and\ \citenamefont {Lisyansky}}]{Perturbation_theory_2020}%
  \BibitemOpen
  \bibfield  {author} {\bibinfo {author} {\bibfnamefont {V.~Y.}\ \bibnamefont {Shishkov}}, \bibinfo {author} {\bibfnamefont {E.~S.}\ \bibnamefont {Andrianov}}, \bibinfo {author} {\bibfnamefont {A.~A.}\ \bibnamefont {Pukhov}}, \bibinfo {author} {\bibfnamefont {A.~P.}\ \bibnamefont {Vinogradov}},\ and\ \bibinfo {author} {\bibfnamefont {A.~A.}\ \bibnamefont {Lisyansky}},\ }\bibfield  {title} {\bibinfo {title} {Perturbation theory for {Lindblad} superoperators for interacting open quantum systems},\ }\href {https://doi.org/10.1103/PhysRevA.102.032207} {\bibfield  {journal} {\bibinfo  {journal} {Phys. Rev. A}\ }\textbf {\bibinfo {volume} {102}},\ \bibinfo {pages} {032207} (\bibinfo {year} {2020})}\BibitemShut {NoStop}%
\bibitem [{\citenamefont {Cirac}\ \emph {et~al.}(1992)\citenamefont {Cirac}, \citenamefont {Blatt}, \citenamefont {Zoller},\ and\ \citenamefont {Phillips}}]{Laser_cooling_perturb_1992}%
  \BibitemOpen
  \bibfield  {author} {\bibinfo {author} {\bibfnamefont {J.~I.}\ \bibnamefont {Cirac}}, \bibinfo {author} {\bibfnamefont {R.}~\bibnamefont {Blatt}}, \bibinfo {author} {\bibfnamefont {P.}~\bibnamefont {Zoller}},\ and\ \bibinfo {author} {\bibfnamefont {W.~D.}\ \bibnamefont {Phillips}},\ }\bibfield  {title} {\bibinfo {title} {Laser cooling of trapped ions in a standing wave},\ }\href {https://doi.org/10.1103/PhysRevA.46.2668} {\bibfield  {journal} {\bibinfo  {journal} {Phys. Rev. A}\ }\textbf {\bibinfo {volume} {46}},\ \bibinfo {pages} {2668} (\bibinfo {year} {1992})}\BibitemShut {NoStop}%
\bibitem [{\citenamefont {Yi}\ \emph {et~al.}(2000)\citenamefont {Yi}, \citenamefont {Li},\ and\ \citenamefont {Su}}]{perturb_dissipation_2000}%
  \BibitemOpen
  \bibfield  {author} {\bibinfo {author} {\bibfnamefont {X.~X.}\ \bibnamefont {Yi}}, \bibinfo {author} {\bibfnamefont {C.}~\bibnamefont {Li}},\ and\ \bibinfo {author} {\bibfnamefont {J.~C.}\ \bibnamefont {Su}},\ }\bibfield  {title} {\bibinfo {title} {Perturbative expansion for the master equation and its applications},\ }\href {https://doi.org/10.1103/PhysRevA.62.013819} {\bibfield  {journal} {\bibinfo  {journal} {Phys. Rev. A}\ }\textbf {\bibinfo {volume} {62}},\ \bibinfo {pages} {013819} (\bibinfo {year} {2000})}\BibitemShut {NoStop}%
\bibitem [{\citenamefont {Andrianov}\ \emph {et~al.}(2020)\citenamefont {Andrianov}, \citenamefont {Ioffe}, \citenamefont {Izotova},\ and\ \citenamefont {Novikov}}]{QME_perturbation_2020}%
  \BibitemOpen
  \bibfield  {author} {\bibinfo {author} {\bibfnamefont {A.~A.}\ \bibnamefont {Andrianov}}, \bibinfo {author} {\bibfnamefont {M.~V.}\ \bibnamefont {Ioffe}}, \bibinfo {author} {\bibfnamefont {E.~A.}\ \bibnamefont {Izotova}},\ and\ \bibinfo {author} {\bibfnamefont {O.~O.}\ \bibnamefont {Novikov}},\ }\bibfield  {title} {\bibinfo {title} {A perturbation algorithm for the pointers of {Franke--Gorini--Kossakowski--Lindblad--Sudarshan} equation},\ }\href {https://doi.org/10.1140/epjp/s13360-020-00540-3} {\bibfield  {journal} {\bibinfo  {journal} {The European Physical Journal Plus}\ }\textbf {\bibinfo {volume} {135}},\ \bibinfo {pages} {531} (\bibinfo {year} {2020})}\BibitemShut {NoStop}%
\bibitem [{\citenamefont {Reiter}\ and\ \citenamefont {S\o{}rensen}(2012)}]{Effective_L_2012}%
  \BibitemOpen
  \bibfield  {author} {\bibinfo {author} {\bibfnamefont {F.}~\bibnamefont {Reiter}}\ and\ \bibinfo {author} {\bibfnamefont {A.~S.}\ \bibnamefont {S\o{}rensen}},\ }\bibfield  {title} {\bibinfo {title} {Effective operator formalism for open quantum systems},\ }\href {https://doi.org/10.1103/PhysRevA.85.032111} {\bibfield  {journal} {\bibinfo  {journal} {Phys. Rev. A}\ }\textbf {\bibinfo {volume} {85}},\ \bibinfo {pages} {032111} (\bibinfo {year} {2012})}\BibitemShut {NoStop}%
\bibitem [{\citenamefont {Kessler}(2012)}]{Effective_L2_2012}%
  \BibitemOpen
  \bibfield  {author} {\bibinfo {author} {\bibfnamefont {E.~M.}\ \bibnamefont {Kessler}},\ }\bibfield  {title} {\bibinfo {title} {Generalized {Schrieffer-Wolff} formalism for dissipative systems},\ }\href {https://doi.org/10.1103/PhysRevA.86.012126} {\bibfield  {journal} {\bibinfo  {journal} {Phys. Rev. A}\ }\textbf {\bibinfo {volume} {86}},\ \bibinfo {pages} {012126} (\bibinfo {year} {2012})}\BibitemShut {NoStop}%
\bibitem [{\citenamefont {Kastoryano}\ \emph {et~al.}(2011)\citenamefont {Kastoryano}, \citenamefont {Reiter},\ and\ \citenamefont {S\o{}rensen}}]{Sorensen_cavity_2011}%
  \BibitemOpen
  \bibfield  {author} {\bibinfo {author} {\bibfnamefont {M.~J.}\ \bibnamefont {Kastoryano}}, \bibinfo {author} {\bibfnamefont {F.}~\bibnamefont {Reiter}},\ and\ \bibinfo {author} {\bibfnamefont {A.~S.}\ \bibnamefont {S\o{}rensen}},\ }\bibfield  {title} {\bibinfo {title} {{Dissipative Preparation of Entanglement in Optical Cavities}},\ }\href {https://doi.org/10.1103/PhysRevLett.106.090502} {\bibfield  {journal} {\bibinfo  {journal} {Phys. Rev. Lett.}\ }\textbf {\bibinfo {volume} {106}},\ \bibinfo {pages} {090502} (\bibinfo {year} {2011})}\BibitemShut {NoStop}%
\bibitem [{\citenamefont {Albert}\ \emph {et~al.}(2016)\citenamefont {Albert}, \citenamefont {Bradlyn}, \citenamefont {Fraas},\ and\ \citenamefont {Jiang}}]{steady_state_theory_2016}%
  \BibitemOpen
  \bibfield  {author} {\bibinfo {author} {\bibfnamefont {V.~V.}\ \bibnamefont {Albert}}, \bibinfo {author} {\bibfnamefont {B.}~\bibnamefont {Bradlyn}}, \bibinfo {author} {\bibfnamefont {M.}~\bibnamefont {Fraas}},\ and\ \bibinfo {author} {\bibfnamefont {L.}~\bibnamefont {Jiang}},\ }\bibfield  {title} {\bibinfo {title} {{Geometry and Response of Lindbladians}},\ }\href {https://doi.org/10.1103/PhysRevX.6.041031} {\bibfield  {journal} {\bibinfo  {journal} {Phys. Rev. X}\ }\textbf {\bibinfo {volume} {6}},\ \bibinfo {pages} {041031} (\bibinfo {year} {2016})}\BibitemShut {NoStop}%
\bibitem [{\citenamefont {Albert}(2018)}]{steady_state_theory_2018}%
  \BibitemOpen
  \bibfield  {author} {\bibinfo {author} {\bibfnamefont {V.~V.}\ \bibnamefont {Albert}},\ }\href {https://arxiv.org/abs/1802.00010} {\bibinfo {title} {Lindbladians with multiple steady states: theory and applications}} (\bibinfo {year} {2018}),\ \Eprint {https://arxiv.org/abs/1802.00010} {arXiv:1802.00010 [quant-ph]} \BibitemShut {NoStop}%
\bibitem [{\citenamefont {Levy}\ \emph {et~al.}(2021)\citenamefont {Levy}, \citenamefont {Rabani},\ and\ \citenamefont {Limmer}}]{response_theory_2021}%
  \BibitemOpen
  \bibfield  {author} {\bibinfo {author} {\bibfnamefont {A.}~\bibnamefont {Levy}}, \bibinfo {author} {\bibfnamefont {E.}~\bibnamefont {Rabani}},\ and\ \bibinfo {author} {\bibfnamefont {D.~T.}\ \bibnamefont {Limmer}},\ }\bibfield  {title} {\bibinfo {title} {Response theory for nonequilibrium steady states of open quantum systems},\ }\href {https://doi.org/10.1103/PhysRevResearch.3.023252} {\bibfield  {journal} {\bibinfo  {journal} {Phys. Rev. Res.}\ }\textbf {\bibinfo {volume} {3}},\ \bibinfo {pages} {023252} (\bibinfo {year} {2021})}\BibitemShut {NoStop}%
\bibitem [{\citenamefont {Li}\ \emph {et~al.}(2014)\citenamefont {Li}, \citenamefont {Petruccione},\ and\ \citenamefont {Koch}}]{perturb_theory_2014}%
  \BibitemOpen
  \bibfield  {author} {\bibinfo {author} {\bibfnamefont {A.~C.~Y.}\ \bibnamefont {Li}}, \bibinfo {author} {\bibfnamefont {F.}~\bibnamefont {Petruccione}},\ and\ \bibinfo {author} {\bibfnamefont {J.}~\bibnamefont {Koch}},\ }\bibfield  {title} {\bibinfo {title} {Perturbative approach to {Markovian} open quantum systems},\ }\bibfield  {journal} {\bibinfo  {journal} {Scientific Reports}\ }\textbf {\bibinfo {volume} {4}},\ \href {https://doi.org/10.1038/srep04887} {10.1038/srep04887} (\bibinfo {year} {2014})\BibitemShut {NoStop}%
\bibitem [{\citenamefont {Li}\ \emph {et~al.}(2016)\citenamefont {Li}, \citenamefont {Petruccione},\ and\ \citenamefont {Koch}}]{Perturbation_infinite_resum_2016}%
  \BibitemOpen
  \bibfield  {author} {\bibinfo {author} {\bibfnamefont {A.~C.~Y.}\ \bibnamefont {Li}}, \bibinfo {author} {\bibfnamefont {F.}~\bibnamefont {Petruccione}},\ and\ \bibinfo {author} {\bibfnamefont {J.}~\bibnamefont {Koch}},\ }\bibfield  {title} {\bibinfo {title} {{Resummation for Nonequilibrium Perturbation Theory and Application to Open Quantum Lattices}},\ }\href {https://doi.org/10.1103/PhysRevX.6.021037} {\bibfield  {journal} {\bibinfo  {journal} {Phys. Rev. X}\ }\textbf {\bibinfo {volume} {6}},\ \bibinfo {pages} {021037} (\bibinfo {year} {2016})}\BibitemShut {NoStop}%
\bibitem [{\citenamefont {Lenarčič}\ \emph {et~al.}(2018)\citenamefont {Lenarčič}, \citenamefont {Lange},\ and\ \citenamefont {Rosch}}]{perturb_Gibbs_2018}%
  \BibitemOpen
  \bibfield  {author} {\bibinfo {author} {\bibfnamefont {Z.}~\bibnamefont {Lenarčič}}, \bibinfo {author} {\bibfnamefont {F.}~\bibnamefont {Lange}},\ and\ \bibinfo {author} {\bibfnamefont {A.}~\bibnamefont {Rosch}},\ }\bibfield  {title} {\bibinfo {title} {Perturbative approach to weakly driven many-particle systems in the presence of approximate conservation laws},\ }\bibfield  {journal} {\bibinfo  {journal} {Physical Review B}\ }\textbf {\bibinfo {volume} {97}},\ \href {https://doi.org/10.1103/physrevb.97.024302} {10.1103/physrevb.97.024302} (\bibinfo {year} {2018})\BibitemShut {NoStop}%
\bibitem [{\citenamefont {Nagib}(2025)}]{nagib_exact_2025}%
  \BibitemOpen
  \bibfield  {author} {\bibinfo {author} {\bibfnamefont {O.}~\bibnamefont {Nagib}},\ }\href@noop {} {\bibinfo {title} {{exact-steady-state-codes}}},\ \bibinfo {howpublished} {\url{https://doi.org/10.5281/zenodo.15750588}} (\bibinfo {year} {2025}),\ \bibinfo {note} {{G}itHub: \url{https://github.com/OmarNagib/exact-steady-state-codes}}\BibitemShut {NoStop}%
\bibitem [{\citenamefont {Happer}\ \emph {et~al.}(2010)\citenamefont {Happer}, \citenamefont {Jau},\ and\ \citenamefont {Walker}}]{optically_pumped_atoms}%
  \BibitemOpen
  \bibfield  {author} {\bibinfo {author} {\bibfnamefont {W.}~\bibnamefont {Happer}}, \bibinfo {author} {\bibfnamefont {Y.-Y.}\ \bibnamefont {Jau}},\ and\ \bibinfo {author} {\bibfnamefont {T.~G.}\ \bibnamefont {Walker}},\ }\bibinfo {title} {{Density Matrix and Liouville Space}},\ in\ \href {https://doi.org/https://doi.org/10.1002/9783527629503.ch4} {\emph {\bibinfo {booktitle} {Optically Pumped Atoms}}}\ (\bibinfo  {publisher} {John Wiley and Sons, Ltd},\ \bibinfo {year} {2010})\ Chap.~\bibinfo {chapter} {4}, pp.\ \bibinfo {pages} {33--47}\BibitemShut {NoStop}%
\bibitem [{\citenamefont {Qojulia}()}]{qojulia_optomech_cooling}%
  \BibitemOpen
  \bibfield  {author} {\bibinfo {author} {\bibnamefont {Qojulia}},\ }\href@noop {} {\bibinfo {title} {{Optomechanical Cavity}}},\ \bibinfo {howpublished} {\url{https://docs.qojulia.org/examples/optomech-cooling/}},\ \bibinfo {note} {accessed 9~July~2025}\BibitemShut {NoStop}%
\bibitem [{\citenamefont {Aspelmeyer}\ \emph {et~al.}(2014)\citenamefont {Aspelmeyer}, \citenamefont {Kippenberg},\ and\ \citenamefont {Marquardt}}]{Cavityoptomechanics}%
  \BibitemOpen
  \bibfield  {author} {\bibinfo {author} {\bibfnamefont {M.}~\bibnamefont {Aspelmeyer}}, \bibinfo {author} {\bibfnamefont {T.~J.}\ \bibnamefont {Kippenberg}},\ and\ \bibinfo {author} {\bibfnamefont {F.}~\bibnamefont {Marquardt}},\ }\bibfield  {title} {\bibinfo {title} {Cavity optomechanics},\ }\href {https://doi.org/10.1103/RevModPhys.86.1391} {\bibfield  {journal} {\bibinfo  {journal} {Rev. Mod. Phys.}\ }\textbf {\bibinfo {volume} {86}},\ \bibinfo {pages} {1391} (\bibinfo {year} {2014})}\BibitemShut {NoStop}%
\bibitem [{\citenamefont {Chan}\ \emph {et~al.}(2011)\citenamefont {Chan}, \citenamefont {Alegre}, \citenamefont {Safavi-Naeini}, \citenamefont {Hill}, \citenamefont {Krause}, \citenamefont {Gröblacher}, \citenamefont {Aspelmeyer},\ and\ \citenamefont {Painter}}]{opto_cool}%
  \BibitemOpen
  \bibfield  {author} {\bibinfo {author} {\bibfnamefont {J.}~\bibnamefont {Chan}}, \bibinfo {author} {\bibfnamefont {T.~P.~M.}\ \bibnamefont {Alegre}}, \bibinfo {author} {\bibfnamefont {A.~H.}\ \bibnamefont {Safavi-Naeini}}, \bibinfo {author} {\bibfnamefont {J.~T.}\ \bibnamefont {Hill}}, \bibinfo {author} {\bibfnamefont {A.}~\bibnamefont {Krause}}, \bibinfo {author} {\bibfnamefont {S.}~\bibnamefont {Gröblacher}}, \bibinfo {author} {\bibfnamefont {M.}~\bibnamefont {Aspelmeyer}},\ and\ \bibinfo {author} {\bibfnamefont {O.}~\bibnamefont {Painter}},\ }\bibfield  {title} {\bibinfo {title} {Laser cooling of a nanomechanical oscillator into its quantum ground state},\ }\href {https://doi.org/10.1038/nature10461} {\bibfield  {journal} {\bibinfo  {journal} {Nature}\ }\textbf {\bibinfo {volume} {478}},\ \bibinfo {pages} {89–92} (\bibinfo {year} {2011})}\BibitemShut {NoStop}%
\bibitem [{Note1()}]{Note1}%
  \BibitemOpen
  \bibinfo {note} {In MATHEMATICA, one can find the nullspace of a Lindbladian \protect \texttt {L} by calling \protect \texttt {Eigenvectors[L, 1, Method -> {"Arnoldi", "Shift" -> 0}]}, and embedding \protect \texttt {L} in the data structure \protect \texttt {SparseArray} to exploit sparsity if present.}\BibitemShut {Stop}%
\bibitem [{\citenamefont {Walker}\ and\ \citenamefont {Happer}(1997)}]{Thad_optica_pump_review}%
  \BibitemOpen
  \bibfield  {author} {\bibinfo {author} {\bibfnamefont {T.~G.}\ \bibnamefont {Walker}}\ and\ \bibinfo {author} {\bibfnamefont {W.}~\bibnamefont {Happer}},\ }\bibfield  {title} {\bibinfo {title} {Spin-exchange optical pumping of noble-gas nuclei},\ }\href {https://doi.org/10.1103/RevModPhys.69.629} {\bibfield  {journal} {\bibinfo  {journal} {Rev. Mod. Phys.}\ }\textbf {\bibinfo {volume} {69}},\ \bibinfo {pages} {629} (\bibinfo {year} {1997})}\BibitemShut {NoStop}%
\bibitem [{\citenamefont {Behary}\ \emph {et~al.}(2024)\citenamefont {Behary}, \citenamefont {Gill}, \citenamefont {Buikema}, \citenamefont {Mikhailov},\ and\ \citenamefont {Novikova}}]{Raman_Ramsey}%
  \BibitemOpen
  \bibfield  {author} {\bibinfo {author} {\bibfnamefont {R.}~\bibnamefont {Behary}}, \bibinfo {author} {\bibfnamefont {A.}~\bibnamefont {Gill}}, \bibinfo {author} {\bibfnamefont {A.}~\bibnamefont {Buikema}}, \bibinfo {author} {\bibfnamefont {E.~E.}\ \bibnamefont {Mikhailov}},\ and\ \bibinfo {author} {\bibfnamefont {I.}~\bibnamefont {Novikova}},\ }\bibfield  {title} {\bibinfo {title} {{Rydberg-Raman-Ramsey resonances in atomic vapor}},\ }\href {https://doi.org/10.1103/PhysRevA.109.053706} {\bibfield  {journal} {\bibinfo  {journal} {Phys. Rev. A}\ }\textbf {\bibinfo {volume} {109}},\ \bibinfo {pages} {053706} (\bibinfo {year} {2024})}\BibitemShut {NoStop}%
\bibitem [{\citenamefont {Rotunno}\ \emph {et~al.}(2023)\citenamefont {Rotunno}, \citenamefont {Robinson}, \citenamefont {Prajapati}, \citenamefont {Berweger}, \citenamefont {Simons}, \citenamefont {Artusio-Glimpse},\ and\ \citenamefont {Holloway}}]{InverseSample}%
  \BibitemOpen
  \bibfield  {author} {\bibinfo {author} {\bibfnamefont {A.~P.}\ \bibnamefont {Rotunno}}, \bibinfo {author} {\bibfnamefont {A.~K.}\ \bibnamefont {Robinson}}, \bibinfo {author} {\bibfnamefont {N.}~\bibnamefont {Prajapati}}, \bibinfo {author} {\bibfnamefont {S.}~\bibnamefont {Berweger}}, \bibinfo {author} {\bibfnamefont {M.~T.}\ \bibnamefont {Simons}}, \bibinfo {author} {\bibfnamefont {A.~B.}\ \bibnamefont {Artusio-Glimpse}},\ and\ \bibinfo {author} {\bibfnamefont {C.~L.}\ \bibnamefont {Holloway}},\ }\bibfield  {title} {\bibinfo {title} {Inverse transform sampling for efficient {Doppler-averaged} spectroscopy simulations},\ }\bibfield  {journal} {\bibinfo  {journal} {AIP Advances}\ }\textbf {\bibinfo {volume} {13}},\ \href {https://doi.org/10.1063/5.0157748} {10.1063/5.0157748} (\bibinfo {year} {2023})\BibitemShut {NoStop}%
\bibitem [{\citenamefont {Miller}\ \emph {et~al.}(2025{\natexlab{a}})\citenamefont {Miller}, \citenamefont {Meyer}, \citenamefont {Montag}, \citenamefont {Nagib}, \citenamefont {Virtanen}, \citenamefont {Elgee},\ and\ \citenamefont {Cox}}]{miller2025rydiqule}%
  \BibitemOpen
  \bibfield  {author} {\bibinfo {author} {\bibfnamefont {B.~N.}\ \bibnamefont {Miller}}, \bibinfo {author} {\bibfnamefont {D.~H.}\ \bibnamefont {Meyer}}, \bibinfo {author} {\bibfnamefont {C.~A.}\ \bibnamefont {Montag}}, \bibinfo {author} {\bibfnamefont {O.}~\bibnamefont {Nagib}}, \bibinfo {author} {\bibfnamefont {T.}~\bibnamefont {Virtanen}}, \bibinfo {author} {\bibfnamefont {P.~K.}\ \bibnamefont {Elgee}},\ and\ \bibinfo {author} {\bibfnamefont {K.~C.}\ \bibnamefont {Cox}},\ }\href {https://github.com/QTC-UMD/rydiqule/tree/joss} {\bibinfo {title} {{RydIQule Version 2}: Enhancing graph-based modeling of {R}ydberg atoms}},\ \bibinfo {howpublished} {Journal of Open Source Software (submitted; review pending)} (\bibinfo {year} {2025}{\natexlab{a}})\BibitemShut {NoStop}%
\bibitem [{\citenamefont {Miller}\ \emph {et~al.}(2025{\natexlab{b}})\citenamefont {Miller}, \citenamefont {Meyer}, \citenamefont {Montag}, \citenamefont {Nagib}, \citenamefont {Virtanen}, \citenamefont {Elgee},\ and\ \citenamefont {Cox}}]{rydiqule_doppler_demo2025}%
  \BibitemOpen
  \bibfield  {author} {\bibinfo {author} {\bibfnamefont {B.~N.}\ \bibnamefont {Miller}}, \bibinfo {author} {\bibfnamefont {D.~H.}\ \bibnamefont {Meyer}}, \bibinfo {author} {\bibfnamefont {C.~A.}\ \bibnamefont {Montag}}, \bibinfo {author} {\bibfnamefont {O.}~\bibnamefont {Nagib}}, \bibinfo {author} {\bibfnamefont {T.}~\bibnamefont {Virtanen}}, \bibinfo {author} {\bibfnamefont {P.~K.}\ \bibnamefont {Elgee}},\ and\ \bibinfo {author} {\bibfnamefont {K.~C.}\ \bibnamefont {Cox}},\ }\href {https://rydiqule.readthedocs.io/en/stable/examples/Analytical%201D%20Doppler%20Demo.html} {\bibinfo {title} {{Analytical 1D Doppler Solver –– Example Notebook, {rydiqule} documentation (v2.1.0)}}},\ \bibinfo {howpublished} {Online documentation} (\bibinfo {year} {2025}{\natexlab{b}}),\ \bibinfo {note} {accessed 16 July 2025}\BibitemShut {NoStop}%
\bibitem [{\citenamefont {Spohn}(1976)}]{steady_state_cond_1976}%
  \BibitemOpen
  \bibfield  {author} {\bibinfo {author} {\bibfnamefont {H.}~\bibnamefont {Spohn}},\ }\bibfield  {title} {\bibinfo {title} {Approach to equilibrium for completely positive dynamical semigroups of {N}-level systems},\ }\href {https://doi.org/https://doi.org/10.1016/0034-4877(76)90040-9} {\bibfield  {journal} {\bibinfo  {journal} {Reports on Mathematical Physics}\ }\textbf {\bibinfo {volume} {10}},\ \bibinfo {pages} {189} (\bibinfo {year} {1976})}\BibitemShut {NoStop}%
\bibitem [{\citenamefont {Kim}\ and\ \citenamefont {Hassler}(2023)}]{L_eigen2023}%
  \BibitemOpen
  \bibfield  {author} {\bibinfo {author} {\bibfnamefont {S.}~\bibnamefont {Kim}}\ and\ \bibinfo {author} {\bibfnamefont {F.}~\bibnamefont {Hassler}},\ }\bibfield  {title} {\bibinfo {title} {Third quantization for bosons: symplectic diagonalization, non-hermitian {Hamiltonian}, and symmetries},\ }\href {https://doi.org/10.1088/1751-8121/acf177} {\bibfield  {journal} {\bibinfo  {journal} {Journal of Physics A: Mathematical and Theoretical}\ }\textbf {\bibinfo {volume} {56}},\ \bibinfo {pages} {385303} (\bibinfo {year} {2023})}\BibitemShut {NoStop}%
\bibitem [{\citenamefont {Ben-Israel}\ and\ \citenamefont {Greville}(2003)}]{benisrael2003generalized}%
  \BibitemOpen
  \bibfield  {author} {\bibinfo {author} {\bibfnamefont {A.}~\bibnamefont {Ben-Israel}}\ and\ \bibinfo {author} {\bibfnamefont {N.~Y.}\ \bibnamefont {Greville}},\ }\href@noop {} {\emph {\bibinfo {title} {Generalized {Inverses}: {Theory} and {Applications}}}},\ \bibinfo {edition} {2nd}\ ed.\ (\bibinfo  {publisher} {Springer},\ \bibinfo {address} {Berlin, Heidelberg},\ \bibinfo {year} {2003})\ Chap.~\bibinfo {chapter} {1}\BibitemShut {NoStop}%
\bibitem [{\citenamefont {Bloch}(1946)}]{Bloch_eq}%
  \BibitemOpen
  \bibfield  {author} {\bibinfo {author} {\bibfnamefont {F.}~\bibnamefont {Bloch}},\ }\bibfield  {title} {\bibinfo {title} {{Nuclear Induction}},\ }\href {https://doi.org/10.1103/PhysRev.70.460} {\bibfield  {journal} {\bibinfo  {journal} {Phys. Rev.}\ }\textbf {\bibinfo {volume} {70}},\ \bibinfo {pages} {460} (\bibinfo {year} {1946})}\BibitemShut {NoStop}%
\bibitem [{\citenamefont {Wang}\ \emph {et~al.}(2021)\citenamefont {Wang}, \citenamefont {Niu},\ and\ \citenamefont {Ye}}]{Bloch_eq2}%
  \BibitemOpen
  \bibfield  {author} {\bibinfo {author} {\bibfnamefont {Y.}~\bibnamefont {Wang}}, \bibinfo {author} {\bibfnamefont {Y.}~\bibnamefont {Niu}},\ and\ \bibinfo {author} {\bibfnamefont {C.}~\bibnamefont {Ye}},\ }\bibfield  {title} {\bibinfo {title} {Optically pumped magnetometer with dynamic common mode magnetic field compensation},\ }\href {https://doi.org/https://doi.org/10.1016/j.sna.2021.113195} {\bibfield  {journal} {\bibinfo  {journal} {Sensors and Actuators A: Physical}\ }\textbf {\bibinfo {volume} {332}},\ \bibinfo {pages} {113195} (\bibinfo {year} {2021})}\BibitemShut {NoStop}%
\end{thebibliography}%

\appendix

\section{Liouville superoperator formalism}\label{appendix:Liouville}
We give a brief overview of the Liouville superoperator formalism used in this present work \cite{optically_pumped_atoms}. A vectorization ${\rm vec}(A)$ maps the operator $A=\sum_{ij} A_{ij} \ket{i} \otimes \bra{j}$ into the following column vector
\begin{equation}\label{vec}
{\rm vec}(A)= \sum_{ij} A_{ij} \ket{i}\otimes\ket{j}
\end{equation}
When applied to the density matrix, $\rho=\sum_{ij} \rho_{ij} \ket{i} \otimes \bra{j}$, it gives the vectorized density matrix
\begin{equation}
\bm{\rho}={\rm vec}(\rho)= \sum_{ij} \rho_{ij} \ket{i}\otimes\ket{j} 
\end{equation}
It is useful to define a Frobenius inner product between two vectorized matrices:
\begin{equation}
{\rm vec}(B)^{\dag}\cdot{\rm vec}(A)={\rm Tr}(B^{\dag}A)
\end{equation}
which can be proved from the vectorization definition [Eq. \eqref{vec}], and the orthonormality of the basis $\ket{i}$ and $\ket{j}$. It follows from the above relation that the expectation value of a Hermitian operator $A$ is given by
\begin{equation}
{\rm Tr}(A \rho)={\rm vec}(A)^{\rm \dag} \cdot \bm{\rho}
\end{equation}
i.e., it is the dot product between the vectorization of $A$ and $\rho$. An important special case is the trace of a density matrix
\begin{equation}
{\rm Tr}(\rho)={\rm Tr}(\mathbb{1} \rho)={\rm vec}(\mathbb{1})^{\rm T} \cdot \bm{\rho}
\end{equation}
where $\rm {A}^T$ denotes the transpose of $A$. Next, we recast the master equation into the superoperator form. Consider the Lindblad master equation \cite{QuantumMasterPro}:
\begin{equation}
\dot{\rho}(t) = -i[H, \rho(t)] + \sum_k \left[ L_k \rho(t) L_k^\dagger - \frac{1}{2} \{ L_k^\dagger L_k, \rho(t) \} \right]
\end{equation}
Using the vectorization identity ${\rm vec}(ABC)=(A \otimes C^{\rm T}){\rm vec}(B)$, this becomes
\begin{equation}\label{QME_appendix}
\dfrac{d\bm{\rho}(t)}{dt}=\mathcal{L}\bm{\rho}(t)
\end{equation}
where the Liouville superoperator $\mathcal{L}$ is given by
\begin{align}
&\mathcal{L}= -i\big(H \otimes\mathbb{1} -\mathbb{1}\otimes H^{\rm T} \big) \\ & \nonumber + \sum_k  \big(L_k \otimes L_k^* - \dfrac{1}{2}\big[L_k^{\dagger} L_k\otimes  \mathbb{1}+ \mathbb{1}  \otimes L_k^{\rm T} L_k^* \big]\big)
\end{align}
For $\mathcal{L}$ to be trace-preserving, it must obey the relation
\begin{equation}\label{trace_perserve}
{\rm vec}(\mathbb{1})^{\rm T} \cdot \mathcal{L}=0
\end{equation}
To see this, take the dot product of ${\rm vec}(\mathbb{1})^{\rm T}$ with the master equation [Eq. \eqref{QME_appendix}]:
\begin{equation}
\dfrac{d}{dt}{\rm vec}(\mathbb{1})^{\rm T} \cdot \bm{\rho}(t)={\rm vec}(\mathbb{1})^{\rm T} \cdot \mathcal{L}\bm{\rho}(t)
\end{equation}
Making the identification ${\rm Tr}(\rho(t))={\rm vec}(\mathbb{1})^{\rm T} \cdot \bm{\rho}(t)$, and imposing the normalization ${\rm Tr}(\rho(t))=1$ at all times and for any $\rho$, the left hand side must vanish for all $t$. This proves that the right-hand side must vanish for any $\mathcal{L}$, i.e., this proves Eq. \eqref{trace_perserve}. An important consequence is that the left nullspace vector $\bm{L}_0$ must be the same, i.e., $\bm{L}_0={\rm vec}(\mathbb{1})$, for all physical Lindbladians $\mathcal{L}$. 
\section{Efficient construction of the Drazin inverse and the propagator}

\subsection{Efficient construction of the Drazin inverse}\label{appnedix:efficient_L0_minus}

\rsub{The Drazin inverse of $\mathcal{L}_0$ is the unique generalized inverse that satisfies the three properties: (a) $(\mathcal{L}_0)^{k+1}\mathcal{L}^-_0=(\mathcal{L}_0)^k$ for some integer $k$, (b) $\mathcal{L}_0\mathcal{L}^-_0=\mathcal{L}^-_0\mathcal{L}_0$, and (c) $\mathcal{L}^-_0\mathcal{L}_0\mathcal{L}^-_0=\mathcal{L}^-_0$.
In the present work we restrict ourselves to a special Drazin inverse that satisfies property (a) for $k=1$, which is called the group inverse. The group inverse has the additional property that (d) $\mathcal{L}_0\mathcal{L}^-_0\mathcal{L}_0=\mathcal{L}_0$ \cite{benisrael2003generalized}.}
 \rsub{If we have the eigendecomposition $\mathcal{L}_0=\sum_{g} g \bm{R}_g \bm{L}_g^T$, then the Drazin inverse satisfying properties (a)-(d) is given by}
 \rsub{\begin{equation}\label{L0_minus_appendix}
\mathcal{L}^-_0=\sum_{g\neq 0}\dfrac{1}{g}\bm{R}_g  \bm{L}_g^T
\end{equation}}
\rsub{It is important to point out that since $\mathcal{L}_0\mathcal{L}^-_0$ and $\mathcal{L}^{-}_0\mathcal{L}_0$ are not Hermitian matrices, $\mathcal{L}^-_0$ is not the Moore-Penrose inverse of $\mathcal{L}_0$  \cite{benisrael2003generalized}}.

\rsub{$\mathcal{L}^-_0$ can be constructed numerically in two different ways as follows. The first approach is based on diagonalization}. Let's denote the right and left eigenvectors matrices of $\mathcal{L}_0$ as $R=(\bm{R}_0,...,\bm{R}_{N-1})$ and $L=(\bm{L}_0,...,\bm{L}_{N-1})$, where the ith column contains the ith eigenvectors, $\bm{R}_i$ and $\bm{L}_i$, respectively. Let $G=(g_0,...,g_{N-1})$ be the vector of eigenvalues $g_i$. $L$ can be directly found by taking the inverse of the transpose of $R$, i.e., $L=(R^{\rm T})^{-1}$. We have observed that in some cases if $R$ is an ill-conditioned matrix, it might be numerically more accurate to compute the inverse operation in $L=(R^{\rm T})^{-1}$ using the Moore-Penrose inverse. The Moore-Penrose inverse is equivalent to the usual matrix inverse if the matrix is invertible. Let $R_-$, $L_-$, and $G_-$ be the matrices and vectors, excluding the nullspaces, i.e., they are $R$ and $L$ without $\bm{R}_0$ and $\bm{L}_0$, and $G_{-}$ is the vector without $g=0$. Then we can succinctly rewrite $\mathcal{L}^-_0$ [Eq. \eqref{L0_minus_appendix}] as 
\begin{equation}\label{L0minus_compact}
\mathcal{L}^-_0=\left(R_{-} * / G_{-} \right) L_{-}^{\rm T}
\end{equation}
where $*/$ denotes the element-wise division of the matrix $R_{-}$ and the vector $G_{-}$, i.e., $(\bm{R}_1/g_1,...,\bm{R}_{N-1}/g_{N-1})$. $( R_{-} * / G_{-} )L_{-}^{\rm T}$ denotes matrix multiplication between $ R_{-} * / G_{-}$ and $L_{-}^{\rm T}$. 

\rsub{The second method of construction avoids diagonalization}
\rsub{\begin{equation}
\mathcal{L}^-_0=(\mathcal{L}_0+\bm{R}_0  \bm{L}_0^T)^{-1}-\bm{R}_0  \bm{L}_0^T
\end{equation}}
\rsub{where $\bm{L}_0={\rm vec}(\mathbb{1})$ for all Lindbladians. Although for a generic $N \times N$ Lindbladian, the cost of constructing the Drazin inverse for both approaches scales as $O(N^3)$, the second method is found to be faster in practice since it only involves a single inverse operation. Note that the second method assumes access to the nullspace vector $\bm{R}_0$, which generally can be obtained faster than a full diagonalization, using efficient sparse linear techniques such as the Arnoldi/Krylov iteration  \cite{Review_2015}.}

\subsection{The \rsub{propagator} for more than one variable}\label{appnedix:L0_minus_generalized}

For the more general problem with several ${v_i}$, i.e.,
\begin{equation}
(\mathcal{L}_0+v_1\mathcal{L}_1+...+v_n \mathcal{L}_n)\bm{\rho}(v_1,v_2,...,v_n)=0
\end{equation}
The corresponding propagator is given by
\begin{equation}\label{Gv_general}
G(v_1,v_2,...,v_n)=\dfrac{1}{\mathbb{1}+\mathcal{L}^-_0  \sum_{i=1}^n v_i\mathcal{L}_i}
\end{equation}
This propagator acts on the zero-case solution $\bm{\rho}_0$ with $\{v_i\}=0$ to generate the general solution $\bm{\rho}(v_1,v_2,...,v_n)$. In the present work, we restrict ourselves to the case of a single $v$.

\subsection{Efficient construction of the propagator}\label{appnedix:efficient_Gv}
$G_v$ is given by Eq. \eqref{Gv}
\begin{equation}\label{Gv_appendix}
G_v=(\mathbb{1}+v\mathcal{L}^-_0 \mathcal{L}_1)^{-1} \equiv \dfrac{\mathbb{1}}{\mathbb{1}+v\mathcal{L}^-_0 \mathcal{L}_1}
\end{equation}
In the eigenbasis of $\mathcal{L}_0^-\mathcal{L}_1$, it is given by Eq. \eqref{Gv_eigen}
\begin{equation}\label{Gv_eigen_appendix}
G_v=\sum_\lambda \dfrac{1}{1+\lambda v} \bm{r}_\lambda \bm{l}^{\rm T}_\lambda 
\end{equation}
Let's denote the right and left eigenvectors matrices of $\mathcal{L}_0^-\mathcal{L}_1$ as $r=(\bm{r}_1,...,\bm{r}_N)$ and $l=(\bm{l}_1,...,\bm{l}_N)$, and the corresponding eigenvalues as the vector $\Lambda=(\lambda_1,...,\lambda_N)$. The left eigenvectors can be computed as $l=(r^{\rm T})^{-1}$. In some problems, if $r$ is an ill-conditioned matrix, it might be numerically more accurate to compute the inverse operation in $l=(r^{\rm T})^{-1}$ using the Moore-Penrose inverse instead of the normal matrix inverse. Using these definitions, $G_v$ can be rewritten compactly as 
\begin{equation}\label{Gv_compact}
G_v=\left[\ r * / (1+ v \Lambda) \right] l^{\rm T}
\end{equation}
where $*/$ denotes the element-wise division of $r$ and $(1+ v \Lambda)$, i.e., $(\bm{r}_1/[1+ v \lambda_1],...,\bm{r}_N/[1+v \lambda_N])$. Note that the scalar $1$ is added to every element of the vector $v \Lambda$ in the denominator. An equivalent form of $G_v$ is:
\rsub{\begin{equation}
G_v=r \ {\rm diag}\left(\dfrac{1}{1+v \Lambda}\right) l^{\rm T}
\end{equation}}
Next, we rewrite $G_v$ further in a way that will prove computationally efficient. Using the identity $1/(1+\lambda v)=1-\lambda v/(1+\lambda v )$ in Eq. \eqref{Gv_eigen_appendix} and the completeness relation $\sum_{\lambda}(\bm{r}_\lambda \bm{l}^{\rm T}_\lambda)=\mathbb{1}$, we can rewrite $G_v$ as:
\begin{equation}\label{Gv_efficient}
G_v=\mathbb{1}-\sum_\lambda \dfrac{\lambda v}{1+\lambda v} \bm{r}_\lambda \bm{l}^{\rm T}_\lambda
\end{equation}
In this form, it is clear that the second term vanishes for all $\lambda=0$, i.e., the nullspace of $\mathcal{L}^-_0 \mathcal{L}_1$ does not contribute to $G_v$. If we denote $r_{-}, \ l_{-},$ and $\Lambda_{-}$ as the matrices and vectors excluding all the zero eigenvectors/values from $r, l,$ and $\Lambda$, then we get:
\begin{equation}\label{Gv_compact_efficient}
G_v= \mathbb{1}- \left[vr_{-}*\Lambda_{-}/ (1+v \Lambda_{-})  \right] l_{-}^{\rm T}
\end{equation}
where $r_{-}*\Lambda_{-}/(1+\Lambda_{-})$ denotes the element wise multiplication of $r_{-}$ and $ \Lambda_{-}/(1+v \Lambda_{-})$. Typically $\mathcal{L}^-_0 \mathcal{L}_1$ has a large nullspace, and so Eq. \eqref{Gv_compact_efficient} is more computationally efficient than unnecessarily including the nullspaces [Eq. \eqref{Gv_compact}].

\section{The relation between the present approach and perturbation theory}\label{appendix:perturbation_theory}

Here, we elaborate on the connection between the present work and perturbation theory of open quantum systems \cite{perturb_theory_2014}. We have shown that the exact solution for an arbitrary $v$ is given by Eq. \eqref{rho_v} 
\begin{equation}\label{rhov_appendix}
\bm{\rho}_v=\dfrac{\mathbb{1}}{\mathbb{1}+v\mathcal{L}^-_0 \mathcal{L}_1}\bm{\rho}_0
\end{equation}
Expanding $\bm{\rho}_v$ in powers of $v$ gives
\begin{equation}\label{rhov_expand}
\bm{\rho}_v=\left[\sum_{n=0}^{\infty}v^n(- \mathcal{L}^-_0 \mathcal{L}_1)^n\right]\bm{\rho}_0
\end{equation}
Truncating this series up to the Mth order gives the perturbative steady state up to the Mth order:
\begin{equation}\label{rhov_perturb}
\bm{\rho}_v^{(M)}=\left[\sum_{n=0}^{M}v^n(- \mathcal{L}^-_0 \mathcal{L}_1)^n\right]\bm{\rho}_0
\end{equation}
Note that even though there is a finite-order truncation, $\bm{\rho}_v^{(M)}$ under the present method is still normalized because $\mathcal{L}_0^-$ preserves the norm. To see this, compute the trace: 
\begin{align}
&{\rm Tr}(\rho_v^{(M)})={\rm vec}(\mathbb{1})^{\rm T}\cdot \bm{\rho}_v^{(M)}= \nonumber \\ & {\rm vec}(\mathbb{1})^{\rm T}(\mathbb{1}-v\mathcal{L}^-_0 \mathcal{L}_1+...+(- v \mathcal{L}^-_0 \mathcal{L}_1)^M)\bm{\rho}_0
\end{align}
The first term is ${\rm vec}(\mathbb{1})^{\rm T}\cdot \bm{\rho}_0=1$, since $\bm{\rho}_0$ is normalized by construction. For higher-order terms, ${\rm vec}(\mathbb{1})^{\rm T} \mathcal{L}_0^-=0$, since ${\rm vec}(\mathbb{1})^{\rm T}=\bm{L}^{\rm T}_0$ is the left nullspace vector for any quantum system, and $\mathcal{L}_0^-$ [Eq. \eqref{L0_minus}] excludes the null space by construction. This shows that $\bm{\rho}_v^{(M)}$ is normalized for any order $M$.  

Eq. \eqref{rhov_perturb} is similar to density-matrix perturbation theory (dubbed Density matrix PT) in Ref. \cite{perturb_theory_2014}, with the difference that the other authors use the Moore-Penrose inverse for their generalized inverse. Density matrix PT suffers from the truncated density matrix $\bm{\rho}_v^{(M)}$ being potentially non-positive for certain parameters \cite{perturb_theory_2014}. In Ref. \cite{Perturbation_infinite_resum_2016}, the authors extend their finite-order perturbation theory, where they do a partial resummation of their perturbative series by decomposing $(-v \mathcal{ L}_0^- \mathcal{L}_1)^n$ into diagonal and off-diagonal parts (in that work they use the same generalized inverse used in this present work $\mathcal{L}_0^-$). They obtain a formal infinite series solution, given by Eq. (31) in their paper, where each term in the series contains corrections up to the infinite order. While Eq. (31) is formally an exact solution, for any practical calculations, this series has to be truncated to finite order, since the series is not in a closed analytic form. 

The power series in Eq. \eqref{rhov_expand} converges if and only if the magnitude of all the eigenvalues of $v\mathcal{L}^-_0 \mathcal{L}_1$ is less than unity, i.e., $|v\mathcal{L}^-_0 \mathcal{L}_1|<1$. In the case where $|v\mathcal{L}^-_0 \mathcal{L}_1|<1$, then a perturbative expansion [Eq. \eqref{rhov_perturb}] exists, which converges to the non-perturbative result [Eq. \eqref{rhov_appendix}] in the limit $M \rightarrow \infty$. If $|v\mathcal{L}^-_0 \mathcal{L}_1|>1$, on the other hand, then a perturbative expansion does not exist. However, even in that case, the non-perturbative result [Eq. \eqref{rhov_appendix}] would still hold exactly for an arbitrary $v$, as was proved in Sec. \ref{sec:propagator_derive}. Thus, this result is exact, non-perturbative, and it works even if a corresponding perturbation theory does not exist. 

\section{Calculation of the Doppler broadening of a two-level system}\label{appendix:doppler}
\begin{figure}[]
   \centering
    \includegraphics[width=0.45\textwidth]{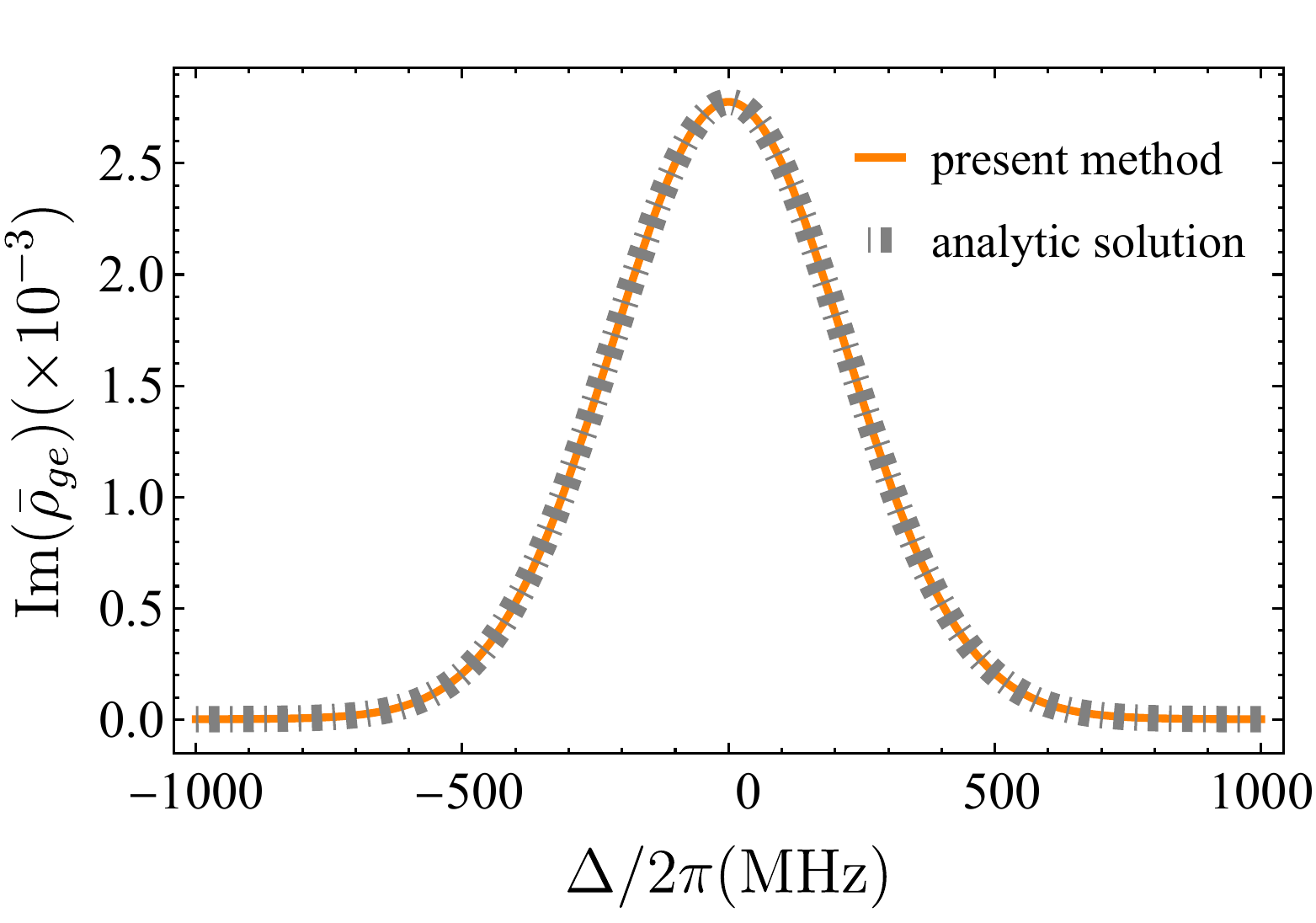}
   \caption{The Doppler broadened absorption spectrum (up to a multiplicative constant) for a two-level system versus the detuning, using the analytic Voigt solution [Eq. \eqref{chi_avg_two_level}] and the present method [Eq. \eqref{rho_gauss_1D}]. The system parameters are $\Gamma= 2\pi \times 6 \ {\rm MHz}$, \ $v_{\rm th}=169.5 \ {\rm \mu m/\mu s}, \ k/2\pi=1/(0.78 \ \rm{\mu m}),$ and $\Omega= 2\pi \times 1 \ {\rm MHz}$.}
    \label{doppler_two_level}
\end{figure}

Consider a two-level system, where an explicit analytic solution exists for the Doppler broadened absorption spectrum. Consider a ground and excited state, $\ket{g}$ and $\ket{e}$, coupled by a field $\Omega$ with a detuning $\Delta$. Let $\Gamma$ be the decay rate of $\ket{e}$. Then the Hamiltonian and the Lindblad operators for this system are given by\begin{subequations}\label{H_two_level}
\begin{align}
&H=-\Delta \ket{e}\bra{e}+\dfrac{\Omega}{2}\ket{e}\bra{g}+\dfrac{\Omega^*}{2}\ket{g}\bra{e}\\ 
&L=\sqrt{\Gamma}\ket{g}\Bra{e}
\end{align}
\end{subequations}
with a corresponding Liouville superoperator 
\begin{equation}
\mathcal{L}_0=\begin{bmatrix}
    -\Gamma & \dfrac{\strut i \Omega}{\strut 2} & -\dfrac{ \strut i \Omega}{\strut 2} & 0 \\
    \dfrac{ \strut i \Omega}{\strut 2} & -\dfrac{ \strut \Gamma}{ \strut 2}+i \Delta & 0 & - \dfrac{i \Omega}{2} \\
    - \dfrac{\strut i \Omega}{2} & 0 & -\dfrac{ \Gamma}{2}-i \Delta &\dfrac{i \Omega}{2} \\
    \Gamma & -\dfrac{i \Omega}{2} & \dfrac{i \Omega}{2} & 0
  \end{bmatrix}
\end{equation}
where $\ket{g}=(0,1)^{\rm T}$ and $\ket{e}=(1,0)^{\rm T}$. The atomic state at steady state $\bm{\rho}_0$ is the nullspace vector of $\mathcal{L}_0$. For this system, it is well-known that the absorption spectrum $\chi$ versus $\Delta$ is a Lorentzian, with its width being determined by the decay rate of the excited state and the intensity of the light field (i.e., power broadening):
\begin{equation}\label{chi_two_level}
\chi \sim \operatorname{Im}(\rho_{\rm ge})=\dfrac{\Omega \Gamma}{\Gamma^2+4\Delta^2+2 \Omega^2}
\end{equation}
For an atom moving at a velocity $v$, the detuning $\Delta$ gets Doppler-shifted as $\Delta-k v$, where $k$ is the wavenumber of the light. An ensemble of atoms moving with different velocities at non-zero temperature obeys a Gaussian distribution, with thermal velocity $v_{\rm th}$. The Doppler-broadened absorption spectrum $\bar{\chi}$ is then calculated by averaging all the $v$-dependent solutions of $\chi$ over that distribution. Applying this procedure on Eq. \eqref{chi_two_level} gives the Doppler-broadened spectrum:
\begin{equation}\label{chi_avg_two_level}
\operatorname{Im}(\bar{\rho}_{\rm ge}) = \dfrac{\pi \Omega \Gamma}{2 \sqrt{\Gamma^2+2 \Omega^2}} V(\Delta;\sqrt{\Gamma^2+2 \Omega^2}/2;k v_{\rm th})
\end{equation}
where $V(.;.;.)$ is the Voigt probability distribution, given by the convolution of a Lorentzian with a half width at half maximum (HWHM) $=\sqrt{\Gamma^2+2 \Omega^2}/2$ and a Gaussian with a standard deviation $=k v_{\rm th}$. 

Using the present approach, we directly obtain the averaged solution by applying Eq. \eqref{rho_gauss_1D}. First, we solve the zero-velocity problem by obtaining $\bm{\rho}_0$ and constructing $\mathcal{L}_0^-$. Next, we diagonalize $\mathcal{L}_0^- \mathcal{L}_1$ to obtain their eigenvalues and eigenvectors. Here, $\mathcal{L}_1$ is the Doppler shifted component of $\mathcal{L}$ given by $\mathcal{L}_1=d\mathcal{L}_0(\Delta-k v)/dv$:
\begin{equation}
\mathcal{L}_1=\begin{bmatrix}
    0 & 0 & 0 & 0 \\
   0 & -i k & 0 & 0 \\
   0 & 0 & ik &0 \\
   0 &0 & 0 & 0
  \end{bmatrix}
\end{equation}
Therefore, the problem of computing Doppler broadening exactly reduces to \rsub{diagonalizating $\mathcal{L}_0^- \mathcal{L}_1$}, as mentioned previously. Fig. \ref{doppler_two_level} shows the Doppler broadened absorption spectrum calculated using the exact Voigt profile and the present method, using system parameters that are typical for $\ce{^{87}Rb}$ atoms. The present method and the analytic solution are in agreement, as expected.

\section{Ensemble average of the steady state over a Lorentzian}\label{appendix:avg}

If $v$ is distributed according to a Lorentzian then we have  $P(v)=\gamma/{\pi(v^2+\gamma^2)}$, where $\gamma$ is the HWHM. This can be integrated analytically to give
\begin{equation}
\bar{\bm{\rho}}=\sum_\lambda f(\lambda,\gamma)\bm{r}_\lambda  (\bm{l}^{\rm T}_\lambda \bm{\rho}_0)
\end{equation}
where 
\begin{equation}
f( \lambda, \gamma) =
\begin{cases}
\dfrac{i}{i + \gamma  \lambda}, & \operatorname{Im}(\lambda) > 0, \\
\dfrac{i}{i - \gamma  \lambda}, &\operatorname{Im}(\lambda) < 0.
\end{cases}
\end{equation}
\section{Magnetometer}

\subsection{Bloch equations steady-state solution}\label{appendix:Bloch}

The Bloch equation of a magnetometer describes the time evolution of the spin $\bm{S}$ as a function of the field $\bm{\Omega}$, optical pumping $\bm{R}$, and spin relaxation $\Gamma$. It is given by \cite{Bloch_eq,Bloch_eq2}
\begin{equation}
\frac{d}{dt}\bm{S} = \bm{\Omega} \times \bm{S} + R \left( \frac{\bm{\hat{R}}}{2} - \bm{S} \right) - \Gamma \bm{S}
\end{equation}
Solving for the steady state gives
\begin{equation}
\bm{S} = \frac{1}{2} \frac{\bm{R} \Gamma' + \bm{\Omega} \times \bm{R} + \bm{\Omega} (\bm{\Omega} \cdot \bm{R})/\Gamma'}{\bm{\Omega}^2 + \Gamma'^2}
\end{equation}
where $\Gamma' = \Gamma + |\bm{R}|$ and we take $\bm{\hat{R}} = \bm{\hat{z}}$ and $ \bm{\Omega} = \Omega_y \bm{\hat{y}} + \Omega_z \bm{\hat{z}}$ in the present calculations.

\subsection{Analytically optimizing the magnetometer response}\label{appenddix:optimize_magnetometer}

We show how to get an analytic expression for $\partial\bm{\rho}(R, \Omega_y=0)/\partial\Omega_y$ as a function of $R$. At steady state, the system as a function of $\Omega_y$ and $R$ is given by 
\begin{equation}
\left(\mathcal{L}_0 + R \mathcal{L}_{R} + \Omega_y \mathcal{L}_{\Omega} \right) \bm{\rho}(R, \Omega_y)=0
\end{equation}
where $\mathcal{L}_\Omega$ is constructed from $S_y$ and $\mathcal{L}_R$ is constructed from the Lindblad operators $S_{+}$ and $S_z$. This has the solution (cf. Appendix \ref{appnedix:L0_minus_generalized}):
\begin{equation}
\bm{\rho}(R, \Omega_y) = \frac{\mathbb{1}}{\mathbb{1} + R \mathcal{L}_0^- \mathcal{L}_{R} + \Omega_y\mathcal{L}_0^- \mathcal{L}_{\Omega}} \bm{\rho}(0, 0)
\end{equation}
Letting $B = \Omega_y \mathcal{L}_0^-\mathcal{L}_{\Omega}$ and $A = R \mathcal{L}_0^-\mathcal{L}_{R}$, we apply the matrix identity 
\begin{equation}
\frac{\mathbb{1}}{\mathbb{1} + A + B} =  \frac{\mathbb{1}}{\mathbb{1} + A} - \frac{\mathbb{1}}{\mathbb{1} + A} B \frac{\mathbb{1}}{\mathbb{1} + A+B}
\end{equation}
on the previous expression. Since we are interested in small $\Omega_y$, we can employ the approximation
\begin{equation}
\frac{\mathbb{1}}{\mathbb{\mathbb{1}} + A + B}  
\approx \frac{\mathbb{1}}{\mathbb{1} + A} - \frac{\mathbb{1}}{\mathbb{1} + A} B \frac{\mathbb{1}}{\mathbb{1} + A}, \ \Omega_y \ll 1
\end{equation}
Applying this identity on $\bm{\rho}(R, \Omega_y)$ gives:
\begin{align}
&\bm{\rho}(R, \Omega_y) \approx \bigg(\frac{\mathbb{1}}{\mathbb{1} + R \mathcal{L}_0^- \mathcal{L}_{R}} \nonumber \\ & - \frac{\mathbb{1}}{\mathbb{1} + R \mathcal{L}_0^- \mathcal{L}_{R}} \Omega_y\mathcal{L}_0^- \mathcal{L}_{\Omega} \frac{\mathbb{1}}{\mathbb{1} + R \mathcal{L}_0^- \mathcal{L}_{R}} \bigg) \bm{\rho}(0, 0)
\end{align}
Taking the derivative with respect to $\Omega_y$ gives the desired result:
\begin{align}
\nonumber &\partial_\Omega \bm{\rho}(R, 0) =  -\frac{\mathbb{1}}{\mathbb{1} + R \mathcal{L}_0^- \mathcal{L}_{R}} \mathcal{L}_0^- \mathcal{L}_{\Omega} \bm{\rho}(R, 0) \\ & = -\frac{\mathbb{1}}{\mathbb{1} + R \mathcal{L}_0^- \mathcal{L}_{R}} \mathcal{L}_0^- \mathcal{L}_{\Omega} \frac{\mathbb{1}}{\mathbb{1} + R \mathcal{L}_0^- \mathcal{L}_{R}} \bm{\rho}(0,0)
\end{align}
This expression can be computed efficiently when expressed in the eigenbasis of $\mathcal{L}_0^- \mathcal{L}_{R}$ (see Appendix \ref{appnedix:efficient_Gv}). Moreover, it can be computed analytically as a function of $R$ for systems sufficiently small in size (e.g., the magnetometer considered presently). Once $\partial_\Omega \bm{\rho}(R, 0)$ is obtained analytically in $R$, it is straightforward to find $R$ that maximizes the magnetometer response.

\end{document}